\documentclass[12pt]{article}      
\usepackage{epsfig,multirow,setspace}

\AtEndOfClass{\RequirePackage{times}}
\usepackage{amssymb, bm}
\usepackage{graphicx}%%%%%%%%%%% standard LaTeX package
\usepackage{comment}
\usepackage{url}
\usepackage{amsmath}
\usepackage{mathtools}
\usepackage{tabularx,booktabs}
\usepackage{float}
\usepackage[font=small,labelfont=bf]{caption}
\usepackage{multirow, makecell}
\usepackage[title]{appendix}
\usepackage{subcaption}
\usepackage{natbib}
\usepackage{hyperref}
\usepackage{longtable}
    \newcolumntype{Y}{>{\centering\arraybackslash}X}
\newcommand{\Bv}{\mathbf{B}}

\newcommand{\Ev}{\mathbf{E}}
\newcommand{\Fv}{\mathbf{F}}

\newcommand{\Iv}{\mathbf{I}}

\newcommand{\Pv}{\mathbf{P}}

\newcommand{\Sv}{\mathbf{S}}

\newcommand{\Vv}{\mathbf{V}}
\newcommand{\Wv}{\mathbf{W}}
\newcommand{\Xv}{\mathbf{X}}
\newcommand{\Yv}{\mathbf{Y}}
\newcommand{\Zv}{\mathbf{Z}}

\newcommand{\ev}{\mathbf{e}}
\newcommand{\fv}{\mathbf{f}}

\newcommand{\xv}{\mathbf{x}}
\newcommand{\yv}{\mathbf{y}}

\usepackage{multirow} % Necessario per \multirow
\usepackage{array} 
\usepackage{color}
\newcommand{\betav}{\mbox{\bm{$\beta$}}}
\newcommand{\epsilonv}{\mbox{\bm{$\epsilon$}}}
\newcommand{\muv}{\mbox{\bm{$\mu$}}}

\newcommand{\Sigmav}{\mbox{\bm{$\Sigma$}}}

\newcommand{\deltav}{\mbox{\bm{$\delta$}}}
\newcommand{\Lambdav}{\mbox{\bm{$\Lambda$}}}

\newcommand{\Omegav}{\mbox{\bm{$\Omega$}}}
\newcommand{\omegav}{\mbox{\bm{$\omega$}}}
\newcommand{\Phiv}{\mbox{\bm{$\Phi$}}}

\newcommand{\thetav}{\mbox{\bm{$\theta$}}}
\newcommand{\piv}{\mbox{\bm{$\pi$}}}
\newcommand{\Gammav}{\mbox{\bm{$\Gamma$}}}

\newcommand{\Psiv}{\mbox{\bm{$\Psi$}}}

\definecolor{brightgreen}{rgb}{0.0, 0.6, 0.0}
\definecolor{patriarch}{rgb}{0.5, 0.0, 0.5}
\newcommand{\zero}{\mathbf{0}}

\newcommand{\half}{\frac{1}{2}}

\DeclareMathOperator\tr{tr}
\title{Extending Cluster-Weighted Factor Analyzers for multivariate prediction and high-dimensional interpretability}
\author{ Xiaoke Qin\footnote{School of Mathematics and Statistics, Carleton University, ON, Canada} \quad Francesca Martella \footnote{Department of Statistical Sciences,  Sapienza University of Rome, Italy} \quad  Sanjeena Subedi \footnotemark[1]}     

\date{}

\begin{document}

\maketitle
\doublespacing
\textbf{Abstract:}
Cluster-weighted factor analyzers (CWFA) are a versatile class of mixture models designed to estimate the joint distribution of a random vector that includes a response variable along with a set of explanatory variables. They are particularly valuable in situations involving high dimensionality.
This paper enhances CWFA models in two notable ways. First, it enables the prediction of multiple response variables while considering their potential interactions. Second, it identifies factors associated with disjoint groups of explanatory variables, thereby improving interpretability.
This development leads to the introduction of the multivariate cluster-weighted disjoint factor analyzers (MCWDFA) model. An alternating expectation-conditional maximization algorithm is employed for parameter estimation. 
The effectiveness of the proposed model is assessed through an extensive simulation study that examines various scenarios.
The proposal is applied to crime data from the United States, sourced from the UCI Machine Learning Repository, with the aim of capturing potential latent heterogeneity within communities and identifying groups of socio-economic features that are similarly associated with factors predicting crime rates.
Results provide valuable insights into the underlying structures influencing crime rates which may potentially be helpful for effective cluster-specific policymaking and social interventions.

\section{Introduction}
\label{sec:1}
Mixture models serve as a powerful statistical approach for clustering observations, which is crucial across various domains, including biology, economics, engineering, and social sciences. \cite{Gershenfeld97} introduced a particular family of Gaussian mixture models known as Cluster-Weighted Models (CWMs),  which \cite{Wedel2002} also referred to as saturated mixture regression models.
For a random vector  $(\Xv, Y)'$, the functional dependence of $Y$ on $\Xv$ is assumed to vary with each mixture component, allowing the component-specific joint density of $(\Xv, Y)'$ to be expressed as  the product of the conditional density of
$Y$ given $\Xv$ and the marginal density of $\Xv$.
\cite{IngrassiaMin2012} reformulated the
CWM within a statistical framework, assuming that both the component-specific conditional distributions of  $Y|\Xv$ and the component-specific marginal distributions of $\Xv$ follow a Gaussian distribution. 
To enhance the applicability of CWMs in high-dimensional $\Xv$-spaces, \cite{SubediPunzo2013} proposed the cluster-weighted factor analyzers (CWFA) model, which addressed this challenge by positing a latent structure for the explanatory variables within each mixture component. In this work, we introduce  a novel mixture model called
the multivariate cluster-weighted disjoint factor analyzers (MCWDFA), which extends the CWFA model in two key aspects. \\
First, the MCWDFA enables the prediction of multiple response variables while accounting for their potential interactions. This leads to a more flexible model that can better capture the complexity and variability inherent in real-world phenomena. By modelling the correlations between the measured variables more precisely, the MCWDFA provides a more comprehensive understanding of the underlying mechanisms governing the case study. \\
Second, the model identifies factors associated with disjoint sets (or groups) of explanatory variables, which similarly predict the responses. This feature simplifies the interpretability of the model by revealing the relationships between distinct groups of predictors and the outcome variables.\\
Specifically, building on the concepts presented by  \cite{Martellaetal2008} and \cite{Vichi2017}, we modify the factor loading matrix by expressing it as the product of a binary row-stochastic matrix and a diagonal matrix within the factor analyzer structure. \\
This approach allows us to group the explanatory variables that similarly predict the responses, ensuring that each explanatory variable has non-zero loading values for only one specific factor. As a result, each variable is exclusively linked to a single factor. This modification not only enhances the interpretability of the resulting factors but also clarifies the interpretation of the numerous regression coefficients, particularly in high-dimensional  $\textbf{X}$-spaces.\\

The paper is structured as follows. In Section \ref{sec:2}, we briefly recall the approach introduced by \cite{SubediPunzo2013}.  Section \ref{sec:3}  illustrates the extension to multivariate responses and disjoint factor analyzers. In Section  \ref{sec:4},  we discuss the alternating expectation-conditional maximization \citep[AECM;][]{Meng1997} algorithm for parameter estimation, along with the convergence criteria and model selection.
Section \ref{sec:6} presents an extensive simulation study in which we evaluate the effectiveness of the proposed approach in accurately recovering the true partitions and model parameter values across various settings and comparing the performance of some information criteria. In Section \ref{sec:7}, we apply the proposed model to a real-world dataset detailing crime events in the United States (sourced from the UCI Machine Learning Repository). After a thorough exploratory analysis, we present the main key findings obtained by the proposed model. Finally, Section \ref{sec:8} offers concluding remarks.

\section{Cluster-weighted factor analyzers model}
\label{sec:2}

The CWFA model \citep{SubediPunzo2013} is a specific type of mixture model designed to estimate the joint distribution of a random vector that includes a response variable and a set of explanatory variables. Within each Gaussian component of the mixture, a single factor analysis regression (FAR) model  \citep{Basilevsky81} is assumed. 
Let $y \in \mathbb{R}$ represent the response variable and $\Xv\in \mathbb{R}^p$  denote the vector of explanatory variables. Together, they form the realizations of the pair ($\Xv, Y$). 
Specifically, the CWFA model posits that:
\begin{equation}
\label{CW}
Y=\beta_{0g}+\betav_{1g}'\Xv+e_g \quad \quad \textrm{with} \quad \quad 
\mathbf{X}=\muv_g+\Lambdav_g\Fv_g+\epsilonv_g,
\end{equation}
with probability $\pi_g$ ($g=1,\dots,G$). The term 
$\muv_g$ represents the component-specific mean vectors of $\Xv$, while $\Lambdav_g$ denotes a $p \times Q$
component-specific factor loadings matrix, where $Q<p$. 

The $\Fv_g$ is a $Q$-dimensional vector of component-specific  factors, assumed to be  i.i.d. draws from a Gaussian distribution $\textrm{MVN}(0, \Iv_Q)$, with  $\Iv_Q$ representing  the $Q \times Q$ identity
matrix. $\epsilonv_g$'s are i.i.d. component-specific errors that follow a Gaussian distribution $\textrm{MVN}(0,\Psiv_g)$, where $\Psiv_g=\textrm{diag}(\psi_{1g}, \dots,\psi_{pg})$,  and these errors are assumed to be independent of $\Fv_g$.
Additionally, $\beta_{0g}$ and $\betav_{1g}$ are the component-specific intercept and the ($1 \times p$) component-specific vector of the regression coefficients, respectively, while $e_g$ is a component-specific disturbance term with a Gaussian distribution $\mathrm{N}(0, \sigma^2_{g})$. 
Furthermore, by assuming that $Y$ is conditionally independent of $\Fv$ given $\Xv=\xv$ in the generic $g$-th  mixture component, we derive  the joint
density of ($\Xv, Y$) as follows:

\begin{equation}
p(\xv,y,\thetav)=\sum_{g=1}^G\pi_g \textrm{N}(y|\xv; m(\xv; \betav_g),\sigma^2_g)\textrm{MVN}(\xv;\muv_g, \Lambdav_g\Lambdav_g'+\Psiv_g),
\end{equation}
where $m(\xv; \betav_g)=\beta_{0g}+\betav_{1g}'\Xv$ and $\thetav=\{
\pi_g,\betav_g, \sigma^2_g,  \muv_g,
\Lambdav_g ,\Psiv_g; g=1,\dots,G\}$. 

Notice that, to promote parsimony, \cite{SubediPunzo2013} impose constraints across groups on $\sigma^2_g$, $\Lambdav_g$, $\Psiv_g$,  
 and on whether $\Psiv_g =\psi_g\Iv_p$ (isotropipc assumption). This approach leads to a collection of sixteen parsimonious CWFA models (see Table 1 in \cite{SubediPunzo2013}). An AECM algorithm can be used for the maximum likelihood estimation of parameters across all models. Further details on AECM initialization and convergence criteria are available in  \cite{SubediPunzo2013}.

\section{Extending CWFA modelling to multivariate responses and disjoint factor analyzers}
\label{sec:3}
To extend the CWFA model discussed in the previous section to accommodate multiple response variables and to identify factors associated with disjoint groups of explanatory variables that similarly predict the responses,   we define, as before, 
$\mathbf{X}$ as  the $p$-dimensional vector of explanatory variables, but  now define $\mathbf{Y}$ as the  $M$-dimensional vector of the response variables. We use the term ``component" to refer to the cluster of the observations and ``segments" to refer to the group of the variables.
For each component $g$ ($g=1,\dots, G$), the proposed MCWDFA model comprises two sub-models. The first modification involves extending the regression model in  (\ref{CW}) to a multivariate regression framework, formalizing the relationships between the $M$ response variables and the  $p$ explanatory variables as follows:
\begin{equation}
\mathbf{Y}=\mathbf{B}_{0g}+\mathbf{B}_{1g}'\mathbf{X}+\mathbf{e}_g.
\end{equation}
Here,  $\mathbf{B}_{0g}$ is the ($M \times 1$) component-specific vector of intercepts,  $\mathbf{B}_{1g}$ is the ($p \times M$) component-specific matrix of regression coefficients, and $\mathbf{e}_g$ is the ($M \times 1$) component-specific vector of disturbances variables, which follows a  Gaussian distribution $\mathrm{MVN}(\mathbf{0}, \Sigma_{\mathbf{e}_g})$.

Furthermore, the second part of the model modifies the factor loading structure of the CWFA model by imposing a structure on the factor loading matrix $\Lambdav_g$. To implement clustering of explanatory variables into disjoint segments that similarly predict the responses, $\Lambdav_g$ is substituted with the product of two specific matrices,  $\mathbf{V}_g$ and $\mathbf{W}_g$. $\mathbf{V}_g=[v_{jqg}]$ is a ($p \times Q$) 
component-specific binary row-stochastic matrix representing the membership of explanatory variables into $Q$ segments corresponding to $Q$ factors.
Specifically, $v_{jqg} = 1$ if and only if, for observations in the $g$-th component,  the $j$-th explanatory variable belongs to segment $q$, and  $v_{jqg} = 0$ otherwise ($j=1,\dots,p$). On the other hand,  $\mathbf{W}_g=\textrm{diag}(w_{1g},\dots,w_{pg})$ is a ($p \times p$) component-specific diagonal matrix of weights for the explanatory variables. The constraint $\mathbf{V}_g'\mathbf{W}_g\mathbf{W}_g\mathbf{V}_g=\textrm{diag}(w_{.1g}^2,\dots,w_{.Qg}^2)$, with $w_{.qg}^2=\sum\limits_j w_{jqg}^2>0$, must be satisfied. Here, 
$w_{jqg}$ refers to the $j$-th variable associated with the $q$-th factor.
Thus, the factor structure in (\ref{CW}) can be constrained  to incorporate the clustering of explanatory variables as follows:
\begin{equation}
\mathbf{X}=\mbox{\boldmath$\mu$}_g+\mathbf{W}_g\mathbf{V}_g\mathbf{F}_g+\mbox{\boldmath$\epsilon$}_g.
\end{equation}

It is worth noting that, within each $g$-th component, an explanatory variable is exclusively associated with a single factor and consequently, the component-specific covariance matrix of $\mathbf{X}$ exhibits a particular structure.
Under assumptions similar to those in the CWFA model, this matrix assumes a block diagonal form, after appropriately permuting the explanatory variables. Each block in this structure represents the component-specific covariance matrix of a subset of explanatory variables related to a specific factor.
 For illustration, suppose we have five explanatory variables clustered into three segments as the following matrices indicate:
  \begin{align}
        \mathbf{W}_g = \begin{pmatrix}
          w_{1g} & 0       &  0 & 0 & 0\\
          0      & w_{2g}  &  0 & 0 & 0\\
          0      & 0       &  w_{3g} & 0 & 0 \\
          0 & 0 & 0 & w_{4g} & 0 \\
          0      & 0       &  0 & 0 & w_{5g}
        \end{pmatrix},
        \hspace{0.5cm}
       \mathbf{V}_g = \begin{pmatrix}
          1 & 0 & 0 \\
          1 & 0 & 0 \\
          0 & 1 & 0 \\
          0 & 1 & 0 \\
          0 & 0 & 1 \\
          \end{pmatrix}.
        \end{align}
 $\mathbf{V}_g$ shows the cluster membership of the five explanatory variables in the three segments while the diagonal entries of $\mathbf{W}_g$ show the weights of explanatory variables on each factor. Assuming 
$\Psiv_g=\textrm{diag}(\psi_{1g}, \psi_{2g},\psi_{3g}, \psi_{4g},\psi_{5g})$, the resulting  component-specific covariance matrix of $\mathbf{X}$ becomes 
 \begin{align*}\small{
        \textbf{W}_g \textbf{V}_g \textbf{V}_g' \textbf{W}_g+ \mbox{\boldmath$\Psi$}_g = \begin{pmatrix}
          w_{1g}^2+\psi_{1g} &  w_{1g}  w_{2g}        &  0 & 0 & 0\\
            w_{1g}  w_{2g}      & w_{2g}^2+\psi_{2g}  &  0 & 0 & 0\\
          0      & 0       &  w_{3g}^2+\psi_{3g} & w_{3g}  w_{4g} & 0 \\
          0 & 0 &  w_{3g}  w_{4g} & w_{4g}^2+\psi_{4g} & 0 \\
          0      & 0       &  0 & 0 & w_{5g}^2+\psi_{5g}
        \end{pmatrix}}
         \end{align*}
where non-zero entries correspond to explanatory variables in the same segment and zero entries for explanatory variables in different segments. Notice that low values of $w_{jg}$ indicate low covariance, thus, different $w_{jg}$ and $\psi_{jg}$ combinations enable each block to capture high/low, low/high, high/high, and low/low variance-correlation patterns.
Thus, recalling that $\mathbf{Y} = \yv$ is conditionally independent of $\mathbf{F}$ given $\mathbf{X}=\mathbf{x}$ within $g$-{th} component, we derive  the joint
density of  $(\Xv, \Yv)$ as follows:

\begin{equation}
 p(\xv, \yv; \mbox{\boldmath$\theta$}) =\sum_{g=1}^{G}\pi_g
\textrm{MVN}({\yv}|\xv; m(\textbf{x}; \widetilde{\mathbf{B}}_g),  {{\Sigmav_{\mathbf{e}_g}}})\textrm{MVN}(\xv; \muv_g, \Wv_g \Vv_g \Vv_g' \Wv_g+ \Psiv_g)
\end{equation}

\noindent where $m(\textbf{x}; \widetilde{\mathbf{B}}_g)=\mathbf{B}_{0g}+\mathbf{B}_{1g}'\mathbf{X}$
with $\widetilde{\mathbf{B}}_g'= \big(\begin{matrix} \mathbf{B}_{0g}  & \mathbf{B}_{1g}' \end{matrix}\big)$  and $\thetav=\{\pi_g,\mathbf{B}_{0g},\mathbf{B}_{1g}, \\ {\Sigmav_{\mathbf{e}_g}},  \muv_g, 
\mathbf{V}_g, \mathbf{W}_g, \Psiv_g\}_{g = 1}^G$ represents the vector of model parameters.  
Note that the MCWDFA model also produces a set of sixteen parsimonious components, akin to those in the CWFA family.

\section{Parameter estimation}
\label{sec:4}

Let $S = \{\yv_i, \xv_i, i=1, \dots, N\}$   represent the set  of all observed data. In a clustering framework, the membership of each observation into different components is unknown. Given $G$ components and a sample size of $N$, an $N\times G$ indicator variable matrix $\Zv$ is employed to capture this membership. Specifically, the entry $z_{ig} = 1$ if the $i$-{th} observation belongs to $g$-th component, and $z_{ig} = 0$ otherwise. Thus, the complete-data likelihood function is defined as:

\begin{align} 
L_c(\thetav)  =\prod_{i = 1}^{N}\prod_{g = 1}^{G}\biggl(
\pi_g & \textrm{MVN} ({\yv}_i|\xv_i, \fv_i; m(\xv_i; \widetilde{\mathbf{B}}_g),  {{\mbox{\boldmath$\Sigma$}_{\mathbf{e}_g}}}) \nonumber\\ 
    & 
\textrm{MVN}
(\xv_i| \fv_{ig}; \mbox{\boldmath$\mu$}_g, \Wv_g\Vv_g\Vv_g'\Wv_g' + \mbox{\boldmath$\Psi$}_g)\textrm{MVN}(\fv_{ig}; \zero, \Iv_{Q})\biggr)^{z_{ig}}
\label{lik}
\end{align} 

\noindent

The idea of the AECM algorithm is to consider a partition of
$\thetav$ in $\thetav_1 = (\pi_g, \mathbf{B}_{0g},\mathbf{B}_{1g}{{\Sigmav_{\mathbf{e}_g}}},\muv_g)_{g = 1}^{G}$ 
and 
$\thetav_2  = ({{\mathbf{W}_g}}, {{\mathbf{V}_g}}, \Psiv_g)_{g = 1}^G$ in such a way that the likelihood is maximized easily for $\thetav_1$  given $\thetav_2$ and vice versa. One iteration consists of two cycles, with one E-step and one CM-step for each cycle. The algorithm iterates the two cycles until the convergence criterion is achieved, yielding the final parameter estimates for $\thetav_1$ and $\thetav_2$. The two cycles at the $(k+1)$-{th} iteration, will be detailed for the more general MCWDFA model in the following section. 

\subsection{First cycle}
In the first cycle, we estimate the  parameters $\thetav_1 = (\pi_g, \mathbf{B}_{0g},\mathbf{B}_{1g},{\Sigmav_{\mathbf{e}_g}}, \muv_g)_{g = 1}^{G}$ consideringonly the indicator variables $z_{ig}$'s as missing data. Then, the complete-data likelihood function is given by
\begin{equation}
    L_{1}(\thetav_1)  = \prod_{i = 1}^{N}\prod_{g = 1}^{G} [\pi_g\textrm{MVN} ({\yv}_i|\xv_i; m(\textbf{x}_i; \widetilde{\mathbf{B}}_g),  {{\Sigmav_{\mathbf{e}_g}}})  \textrm{MVN}
(\xv_i; \muv_g, \Wv_g \Vv_g \Vv_g' \Wv_g+ \Psiv_g)]^{z_{ig}},
\label{li1}
\end{equation}
and the consequent complete-data log-likelihood is

\begin{align}
    \ell_{1}(\thetav_1) & = \sum_{i = 1}^{N}\sum_{g = 1}^{G} z_{ig}[\log\pi_g + \log\textrm{MVN} ({\yv}_i|\xv_i; m(\textbf{x}_i; \widetilde{\mathbf{B}}_g),  {{\Sigmav_{\mathbf{e}_g}}}) + \log \textrm{MVN}
(\xv_i; \muv_g, \Gammav_g)]\nonumber\\ \nonumber
    & = -\frac{N(M+p)}{2}\log(2\pi)- \frac{1}{2}\sum_{i = 1}^{N}\sum_{g = 1}^{G} z_{ig}\log|{{\Sigmav_{\mathbf{e}_g}}}|\\ \nonumber
    &\quad 
     -\frac{1}{2} \sum_{i = 1}^{N}\sum_{g = 1}^{G} z_{ig}  (\yv_i -  \mathbf{B}_{0g}-\mathbf{B}_{1g}\xv_i)'{{\Sigmav_{\mathbf{e}_g}^{-1}}}(\yv_i - \mathbf{B}_{0g}-\mathbf{B}_{1g}\xv_i)  \\ \nonumber
    &\quad - \half \sum_{i = 1}^{N}\sum_{g = 1}^{G} z_{ig} \log| \Gammav_g| - \half\sum_{i = 1}^{N}\sum_{g = 1}^{G} z_{ig}(\xv_i - \muv_g)' \Gammav_g^{-1}(\xv_i - \muv_g) \\ \nonumber
    &\quad + \sum_{i = 1}^{N}\sum_{g = 1}^{G}z_{ig} \log \pi_g ,\\
    \label{log1}
\end{align}
where $ \Gammav_g = \Wv_g\Vv_g\Vv_g'\Wv_g +  \Psiv_g$ for $g = 1,\dots, G$.\\
The E-step of this cycle consists of calculating $Q_1(\thetav_1|\thetav^{(k)})=\mathbb{E}_{\thetav^{(k)}}(\ell_1(\thetav_1)|S)$ which requires the computation of $\mathbb{E}_{\thetav^{(k)}}(z_{ig}|S)$ in practice. Hence, we update $\mathbb{E}_{\thetav^{(k)}}(z_{ig}|S)$ as 
\begin{equation*}
    \mathbb{E}_{\thetav^{(k)}}(z_{ig}|S)=z_{ig}^{(k+1)} = \frac{\pi_g^{(k)}\textrm{MVN} ({\yv}_i|\xv_i; m(\textbf{x}_i; \widetilde{\mathbf{B}}_g^{(k)}),  {{\Sigmav_{\mathbf{e}_g}^{(k)}}})  \textrm{MVN}(\xv_i; \muv_g^{(k)},\Gammav_g^{(k)})} {\sum_{j = 1}^{G}\pi_j^{(k)}\textrm{MVN} ({\yv}_i|\xv_i; m(\textbf{x}_i; \widetilde{\mathbf{B}}_j^{(k)}),  {{\Sigmav_{\mathbf{e}_j}^{(k)}}})  \textrm{MVN}(\xv_i; \muv_j^{(k)},\Gammav_j^{(k)})}.
\end{equation*}
In the M-step, we maximize the objective function (\ref{log1}) obtaining the following estimation
\begin{align}
    &\pi_g^{(k+1)} = \frac{1}{N}\sum_{i = 1}^{N}z_{ig}^{(k+1)}\\
    &\muv_g^{(k+1)} = \frac{1}{N_g^{(k+1)}}\sum_{i = 1}^{N}z_{ig}^{(k+1)}\xv_i\\
    &\widetilde\Bv_g'^{(k+1)} = (\sum_{i = 1}^{N}z_{ig}^{(k+1)}\yv_i\widetilde\xv_i')(\sum_{i=1}^{N} z_{ig}^{(k+1)}\widetilde\xv_i\widetilde\xv_i')^{-1}\\
    & \Sigmav_{\mathbf{e}_j}^{(k+1)}= \frac{1}{N_g^{(k+1)}}\sum_{i = 1}^{N}z_{ig}^{(k+1)}(\yv_i - \widetilde\Bv_{g}'^{(k+1)}\widetilde\xv_i)(\yv_i - \widetilde\Bv_{g}'^{(k+1)}\widetilde\xv_i)',
\end{align}
where $\widetilde\xv_i=[1 \quad \xv_i]$ and $N_g^{(k+1)} = \sum_{i=1}^{N}z_{ig}^{(k+1)}$. The updated parameter is denoted by $\thetav^{(k+\half)} = (\thetav_1^{(k+1)}, \thetav_2^{(k)})$.

\subsection{Second cycle}
In the second cycle, we update $\thetav_2 = (\Wv_g, \Vv_g, \Psi_g)_{g = 1}^{G}$, and both $\fv_{ig}$'s and  $z_{ig}$'s are treated as unobserved data. The complete-data likelihood function is 

\begin{align*}
L_2(\thetav_2) 
           = \prod_{i = 1}^{N}\prod_{g = 1}^{G}[\pi_g^{(k+1)}
& \textrm{MVN} ({\yv}_i|\xv_i, \fv_{ig}; m(\xv_i; \widetilde{\mathbf{B}}_g^{(k+1)}),  {{\Sigmav_{\mathbf{e}_g}^{(k+1)}}}) \\
& \textrm{MVN} (\xv_i|\fv_{ig}; \muv_g^{(k+1)}, \Wv_g\Vv_g\Vv_g'\Wv_g' +  \Psiv_g) \textrm{MVN}(\fv_{ig}; \zero, \Iv_{Q})]^{z_{ig}},
\end{align*}
and thus the complete-data log-likelihood function is 
\begin{align*}
\ell_{2}(\thetav_2) & = -\frac{N(M+J+Q)}{2}\log(2\pi) + \sum_{i = 1}^{N}\sum_{g = 1}^{G} z_{ig}^{(k+1)}\log\pi_g^{(k+1)}\\
    &\quad - \half\sum_{i = 1}^{N}\sum_{g = 1}^{G} z_{ig}\log|\Psiv_g| \\
    &\quad - \half\sum_{i = 1}^{N}\sum_{g = 1}^{G} z_{ig}\tr\biggl((\xv_i - \muv_g^{(k+1)}-\mathbf{W}_g\mathbf{V}_g\fv_{ig})(\xv_i - \muv_g^{(k+1)}-\mathbf{W}_g\mathbf{V}_g\fv_{ig})'\Psiv_g^{-1}\biggr) \\  
    &\quad - \half  \sum_{i = 1}^{N}\sum_{g = 1}^{G} z_{ig}  (\yv_i -  \mathbf{B}_{0g}^{(k+1)}-\mathbf{B}_{1g}^{(k+1)}\xv_i)'(\Sigmav_{\mathbf{e}_g}^{(k+1)})^{-1}(\yv_i - \mathbf{B}_{0g}^{(k+1)}-\mathbf{B}_{1g}^{(k+1)}\xv_i) \\  
    &\quad 
    - \frac{1}{2}\sum_{i = 1}^{N}\sum_{g = 1}^{G} z_{ig}\log|{{\Sigmav_{\mathbf{e}_g}}}|.
\end{align*}

To obtain the expected value $Q_2(\thetav_2;\thetav^{(k+1/2)}) = \mathbb{E}_{\thetav^{(k+1/2)}}(\ell_2(\thetav_2)|S)$, we need to calculate $\mathbb{E}_{\thetav^{(k+1/2)}}(z_{ig}|S)$,
$\mathbb{E}_{\thetav^{(k+1/2)}}(z_{ig}\fv_{ig}|S)$ and $\mathbb{E}_{\thetav^{(k+1/2)}}(z_{ig}\fv_{ig}\fv_{ig}'|S)$. 
Denoting $\deltav_g =  \Vv_g'\Wv_g'(\Wv_g\Vv_g\Vv_g'\Wv_g'+\Psiv_g)^{-1}$, they are given by
\begin{equation*}
   \mathbb{E}_{\thetav^{(k+1/2)}}(z_{ig}\fv_{ig}|S) = z_{ig}^{(k+1)}\deltav_g^{(k)}(\xv_i - \muv_g^{(k+1)}),
\end{equation*}
\begin{equation*}
    \mathbb{E}_{\thetav^{(k+1/2)}}(z_{ig}\fv_{ig}\fv_{ig}'|S) =z_{ig}^{(k+1)}\biggl( \Iv_Q - \deltav_g^{(k)}\Wv_g^{(k)}\Vv_g^{(k)} + \deltav_g^{(k)}\Sv_{g}^{(k+1)}\deltav_g'^{(k)}\biggr)=z_{ig}^{(k+1)}\omegav_g^{(k)} 
\end{equation*}
where $$\Sv_{g}^{(k+1)} = \frac{1}{N_g^{(k+1)}}\sum_{i = 1}^{n}z_{ig}^{(k+1)}(\xv_i - \muv_g^{(k+1)})(\xv_i - \muv_g^{(k+1)})'$$
and $$\omegav_g^{(k)}  = \Iv_Q - \deltav_g^{(k)}\Wv_g^{(k)}\Vv_g^{(k)} + \deltav_g^{(k)}\Sv_g^{(k+1)}\deltav_g'^{(k)}.$$

Therefore, omitting those terms unrelated to $\mbox{\boldmath$\theta$}_2$, the expected complete-data log-likelihood function $Q_2(\mbox{\boldmath$\theta$}_2;\mbox{\boldmath$\theta$}^{(k+1/2)})$ is given by

\begin{align*}
Q_2(\Wv_g,\Vv_g, \Psiv_g   ;\thetav^{(k+1/2)}) &= 
    - \half\sum_{i = 1}^{N} z_{ig}\log|\Psiv_g| \\
    &\quad - \half\sum_{i = 1}^{N} z_{ig}\tr\biggl(\Sv_{g}^{(k+1)}\Psiv_g^{-1}\biggr) \\  
    &\quad + \sum_{i = 1}^{N} z_{ig}\tr\biggl(\Wv_g\Vv_g\deltav^{(k)}\Sv_{g}^{(k+1)}
 \Psiv_g^{-1}\biggr)\\  
    &\quad -\half \sum_{i = 1}^{N} z_{ig} \tr\biggl( (\Wv_g\Vv_g)'\Psiv_g^{-1} (\Wv_g\Vv_g)\omegav_g^{(k)} \biggr)
\end{align*}

The maximization of $Q_2$ with respect to $\Wv_g$ and  $\mbox{\boldmath$\Psi$}_g$ respectively yields:
\begin{gather*}
(\Wv_g)_{jj}^{(k+1)} = (\Sv_{g}^{(k+1)}\deltav'^{(k)}\Vv_{g}')_{jj} / (\Vv_{g}\omegav^{(k)}\Vv_{g}')_{jj}^{(k+1)}, \\
(\Psiv_{g})_{jj}^{(k+1)} = (\Sv_g^{(k+1)} - 2\Sv_g^{(k+1)}\deltav_g'^{(k)}\Vv_g'\Wv_g' + \Wv_g\Vv_g\omegav_g^{(k)}\Vv_g'\Wv_g')_{jj}.
\end{gather*}
In order to estimate the segment membership matrix $\Vv_g$, we proceed as follows:
\begin{itemize}
\item For each explicative variable $j$ and segment $q$, compute the log-likelihood values 
\begin{align*}
    l_{jq}=l(\cdot,v_{jq}=1|& \thetav_1^{(k+1)}, \{\muv_g^{(k+1)}, 
 \mathbf{W}_g^{(k+1)}, \Psiv_g^{(k+1)}\}_{g = 1}^G,\{v_{hs}^{(k+1)}\}_{h < j}, \{v_{hs}^{(k)}\}_{h>j}).
\end{align*}
\item Fix $j$ and compute the maximum of this set $\{l_{jq}\}$ over $q=1,\dots,Q$; denote this term by $l_j^{max}$;
\item Allocate the $j$-th explicative variable to the $q$-{th} segment ($\hat{v}_{jq}=1$) iff $l_{jq}=l_j^{max}$ $q=1,\dots,Q$.
\end{itemize}

In summary, the AECM algorithm consists of the following steps. Denoting $\thetav^{(0)}$ as the  initial value for $\thetav$, on the $(k+1)$-st iteration, it updates $\thetav$ by alternately performing the following steps for $g=1,\dots, G$:

\begin{enumerate}
\item Update $z_{ig}^{(k+1)}$
\item  Update   $\pi_g^{(k+1)},\muv_g^{(k+1)}, \widetilde\Bv_g'^{(k+1)}, \Sigmav_{\mathbf{e}_{g}}^{(k+1)}$;
\item Update 
$\mathbb{E}_{\thetav^{(k+1/2)}}(z_{ig}\fv_{ig}|S)$ and $\mathbb{E}_{\mbox{\boldmath$\theta$}^{(k+1/2)}}(z_{ig}\fv_{ig}\fv_{ig}'|S)$ 
\item Update  $\Vv_g^{(k+1)}$,  $\Wv_g^{(k+1)}$ and $\Psiv_g^{(k+1)}$.
\end{enumerate}

At convergence, each observation is allocated to the component with the highest posterior probability $\hat{z}_{ig}$ (following the MAP rule). At the same time, each explanatory variable is assigned to the $q$-{th} segment based on the entries in the matrix  $\Vv_g$ ($g=1,\dots, G$). Despite numerous proposed strategies for selecting reasonable starting points, no single approach consistently outperforms the others. For a comprehensive overview, refer to \cite{giordani2020introduction}.

\subsection{Convergence criterion and model selection}
Regarding the stopping criterion, we use Aitken's acceleration procedure \citep{Aitken} to determine when the algorithm should terminate.
 This criterion estimates the maximum of the incomplete log-likelihood function as follows:
\begin{align}\label{aac:est}
	\ell_A^{(k)} = \ell^{(k)} + \frac{1}{1 - c^{(k)}}(\ell^{(k+1)} - \ell^{(k)}), 
\end{align}
where $$c^{(k)} = \frac{\ell^{(k+1)} - \ell^{(k)}}{\ell^{(k)} - \ell^{(k-1)}}$$
and 
$\ell^{(k-1)}, \ell^{(k)}, \ell^{(k+1)}$ are updated values of the incomplete log-likelihood function from iterations $k-1$, $k$ and $k+1$. The algorithm stops when $l_A^{(k+1)} - l^{(k)} <\epsilon$ where $\epsilon$ is a small enough positive number.

Lastly, it should be noted that while the AECM algorithm assumes that the number of mixture components and segments for the explanatory variables (i.e. factors) are known, in practice, these quantities must be estimated from the data. As in standard mixture models, we can make use of information criteria.

Here, we consider the following common criteria:
\begin{align*}
    \text{AIC}(\hat{\mbox{\boldmath$\theta$}}) &= -2\ell(\hat{\mbox{\boldmath$\theta$}})+2\eta,\\
		    \text{AIC3}(\hat{\mbox{\boldmath$\theta$}}) &= -2\ell(\hat{\mbox{\boldmath$\theta$}})+3\eta,\\
		    \text{BIC}(\hat{\mbox{\boldmath$\theta$}}) &= -2\ell(\hat{\mbox{\boldmath$\theta$}})+\eta\log(N),\\
		   \text{ICL}(\hat{\mbox{\boldmath$\theta$}}) &= BIC(\hat{\mbox{\boldmath$\theta$}}) - \sum_{i=1}^N\sum_{g=1}^G\hat z_{ig}\log(\hat z_{ig})
\end{align*}
Here, $\ell(\cdot)$ represents the log-likelihood value, and $\eta$ is the number of parameters. A model with a higher value of these criteria is preferable to others.

\section{Simulation Studies}
\label{sec:6}

\subsection{Simulation 1}
In the first simulation study,  our primary goals were to investigate impact of having segments of explanatory variables with either equal or different classifications for each component, and to assess how the number of components $G$ affects model performance. Additionally, we evaluated the effectiveness of various information criteria in identifying the correct number of components $G$ and segments $Q$.
To achieve these objectives, we considered several different settings as follows:

\begin{itemize}
\item \textbf{Setting 1}: $G=2$ with $\piv = (0.6, 0.4)$, $Q=3$ segments of explanatory variables of  equal size 
\item \textbf{Setting 2}:   $G=2$ with $\piv = (0.6, 0.4)$,  $Q=3$ segments of explanatory variables with a different classification for each component 
\item \textbf{Setting 3}:  $G=3$ with  $\piv = (0.37,0.33,0.3)$, $Q=3$ segments of explanatory variables with a different classification for each component.
\end{itemize}
The parameters  $\muv_g$, $\mathbf{B}_{0g}$, $\mathbf{B}_{1g}$, $\Sigmav_{\mathbf{e}_g}$, $\mathbf{W}_g$, $\mathbf{V}_g$ and $\Psiv_{_g}$ for $g = 1, \dots, G$  were pre-specified and are detailed in Tables \ref{simu1:setting1:parameters}, \ref{simu1:setting2:parameters} and \ref{simu1:setting3:parameters} in Appendix \ref{A}, corresponding to Setting 1, 2 and 3, respectively. For each setting, we simulated 100 datasets and ran the AECM algorithm described in Section \ref{sec:4} on each dataset for $G = 1, \dots, 5$ and $Q = 1,\dots, 5$.\\
At the starting point of each run, the initial clustering $\Zv^{(0)}$ and sizes of components $N_g^{(0)}$ were obtained from a $k$-means algorithm. The loading matrix $\Lambdav_g^{(0)} = \mathbf{W}_g^{(0)} \mathbf{V}_g^{(0)}$ was obtained according to a principle component analysis. We can further have the elements of $\mathbf{V}_g^{(0)}$
\begin{align*}
    (\mathbf{V}_g^{(0)})_{jq} = 
    \begin{cases}
        1, \quad \Lambdav^{(0)}_{jq} = \text{max}\{|\Lambdav^{(0)}_{j1}|, |\Lambdav^{(0)}_{j2}|, \dots, |\Lambdav^{(0)}_{jQ}|\}\\
        0, \quad \text{otherwise.}
    \end{cases}
\end{align*}
and the diagonal elements of $\Wv_{g}^{(0)}$ were 
\begin{align*}
    (\mathbf{W}_g^{(0)})_{jj} =  \text{max}\{|\Lambdav^{(0)}_{j1}|, |\Lambdav^{(0)}_{j2}|, \dots, |\Lambdav^{(0)}_{jQ}|\}.
\end{align*}
Let $\Xv_{g}^{(0)}$ be the explanatory variables in $g$-th segment at the starting point, and we can have the estimated covariance as $\Omegav_{g}^{(0)} = \frac{1}{N_g^{(0)} - 1}\Xv_{g}^{(0)}\Xv_{g}^{'(0)}$. Then, the initial elements of $\Psiv_{g}^{(0)}$ were

\begin{align*}
    (\Psiv_{g}^{(0)})_{jj} = |(\Omegav_{g}^{(0)} - \Wv_{g}^{(0)}\Vv_{g}^{(0)}\Vv_{g}^{'(0)}\Wv_{g}^{(0)})_{jj}|.
\end{align*}

\subsubsection{Performance assessment}
To evaluate the performance of the proposed approach in accurately recovering
the true partitions and model parameter values and to  compare the performance of the several information criteria in selecting the correct number of $G$ and $Q$ for our model, we employed the following:
\begin{itemize}
\item[(i)] \textbf{Average Adjusted Rand Index} (ARI, \cite{Hubert1985}) for the unit-specific partition across the 100 simulated datasets. ARI measures the agreement between the true and estimated unit-specific partitions, adjusting for chance. An ARI score of $0$ indicates random labelling, while a score of 1 indicates perfect agreement between the true and estimated partitions; 
\item[(ii)] \textbf{Clustering of Explanatory Variables}: We compared the estimated cluster-specific correlation matrices of $\Xv$ with the true cluster-specific correlation matrices $\Xv$ using heatmaps.  The darkness of the regions in these heatmaps reflects the frequency of non-zero entries across the 100 simulated datasets.
\item[(iii)] \textbf{Frequency of True Model Selection}: We recorded how often the true model was selected by AIC, AIC3, BIC, and ICL across 100 simulated datasets.  This measure allows us to compare the effectiveness of different information criteria in identifying the correct number of components $G$ and cluster $Q$; 
\item[(iv)] The estimated \textbf{Mean Squared Error (MSE)} of model parameters when the true model (by using ICL) was selected.  It is obtained as the average squared differences between the estimated and the true parameter values across the 100 simulated datasets. A smaller MSE indicates better model performance in parameter estimation.
\end{itemize}

\subsubsection{Simulation 1 Results}
Tables \ref{simu1:model_selection} and \ref{simu1:model_ARI} present the results of our model selection criteria and clustering accuracy for Settings 1, 2, and 3, with the true model for each setting highlighted in bold.
Specifically, Table \ref{simu1:model_selection} illustrates the frequency with which each model was chosen based on different criteria (AIC, AIC3, BIC, ICL) for various combinations of components $G$ and segments $Q$. Table \ref{simu1:model_ARI} provides the average ARI values for the selected models.
The criteria generally perform well in identifying the correct number of components $G$ and segments $Q$ for Settings 1 and 2. In Setting 3, the criteria exhibit slightly better performance, with the correct model being selected 100 times for three of the criteria. Additionally, the average ARI reaches 1 across all settings when the correct model is chosen.

\begin{table}[H]
\centering
\scriptsize
\begin{tabular}{cccccc}
\toprule
&\multicolumn{5}{c}{\textbf{Times each  model was selected by (AIC, AIC3, BIC, ICL)}} \\
\hline
\multicolumn{5}{l}{{\textbf{Setting 1}}} \\
\hline
&$G$ = 1 & $G$ = 2 & $G$ = 3 & $G$ = 4 & $G$ = 5 \\
\hline
$Q$ = 1 & - & - & - & - & -  \\
$Q$ = 2 & - & - & - & - & -  \\
$Q$ = 3  & - & {{\textbf{(88,91,96,96)}}} & (4,2,2,2) &(7,5,1,1) & (1,1,0,0)  \\
$Q$ = 4 & - & - & (0,1,1,1) & - & -  \\
$Q$ = 5 & - & - & - & - & -  \\
\hline
\hline
\multicolumn{5}{l}{{\textbf{Setting 2}}} \\
\hline
$Q$ = 1 & - & - & - & - & -   \\
$Q$ = 2 & - & - & - & - & -   \\
$Q$= 3 & - &  {\textbf{(91,93,99,99)}}& (1,1,0,0)& (5,4,0,0) & (2,2,1,1)  \\
$Q$ = 4 & - & - & - & - & (1,0,0,0) \\
$Q$ = 5 & - & - & - & - & -  \\
\hline
\hline
\multicolumn{5}{l}{{\textbf{Setting 3}}} \\
\hline
$Q$ = 1 & - & - & - & - & -   \\
$Q$ = 2 & - & - & - & - & -   \\ 
$Q$ = 3 & - & - &  {\textbf{(97,100,100,100)}}& (3,0,0,0)  & - \\
$Q$ = 4 & - & - & - & - & -  \\
$Q$ = 5 & - & - & - & - & - \\
\hline
\end{tabular}
\caption{Selection frequency of model criteria  for Settings 1, 2 and 3.}
\label{simu1:model_selection}
\end{table}

\begin{table}[H]
\centering
\scriptsize
\begin{tabular}{ccccc}
\toprule
&\multicolumn{4}{c}{\textbf{Average ARI when the model was selected}}\\
\hline
\multicolumn{4}{l}{{\textbf{Setting 1}}} \\
\hline
& $G$ = 2 & $G$ = 3 & $G$ = 4 & $G$ = 5\\
\hline
$Q$ = 1 &  - & - & - &-\\
$Q$ = 2 & - & - & - &-\\
$Q$ = 3  & {\textbf{(1,1,1,1)}} & (0.78, 0.78, 0.78, 0.78) & (0.66,0.67,0.76,0.76)&(0.43,0.43,-,-)\\
$Q$ = 4 &  - & (-,0.81,0.81,0.81) & - &-\\
$Q$ = 5 & - & - & - & - \\
\hline
\hline
\multicolumn{4}{l}{{\textbf{Setting 2}}} \\
\hline
$Q$ = 1 &  - & - & - &- \\
$Q$ = 2 &  - & - & - &- \\
$Q$= 3 &  {\textbf{(1,1,1,1)}} &  (0.75,0.83,-,-)  &  (0.64,0.61,-,-) & (0.49,0.49,0.52,0.52) \\
$Q$ = 4 & - & - & - & (0.46,-,-,-)\\
$Q$ = 5 & - & - & - & -   \\
\hline
\multicolumn{4}{l}{{\textbf{Setting 3}}} \\
\hline
$Q$ = 1 & - & - & - &-  \\
$Q$ = 2 &  - & - & - &- \\ 
$Q$ = 3 &  -  &{\textbf{(1,1,1,1)}} &   (0.86, -, -, -) & -\\
$Q$ = 4 &  - & - & - &-\\
$Q$ = 5 & - & - & - &-\\
\hline
\end{tabular}
\caption{ARI values for Settings 1, 2 and 3.}
\label{simu1:model_ARI}
\end{table}

\begin{figure}
\centering
\begin{subfigure}{.7\textwidth}
\centering
\includegraphics[width=\textwidth]{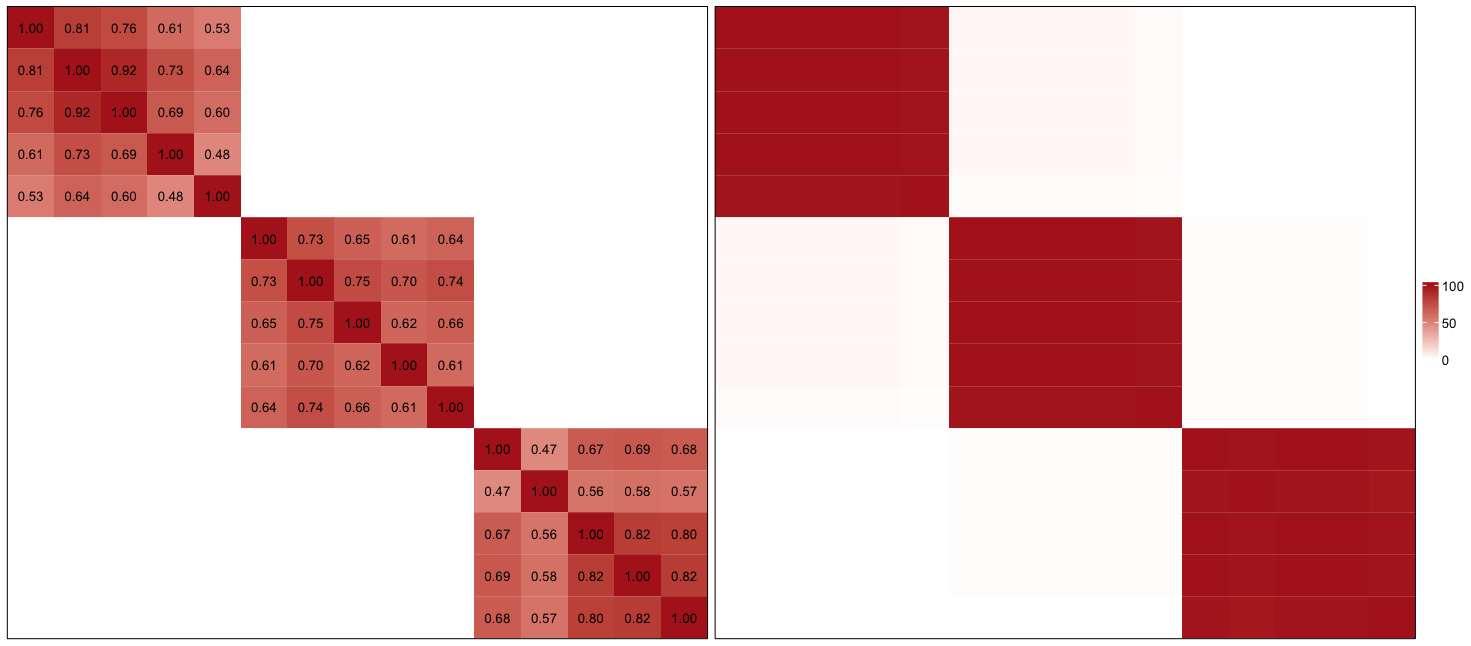}
\caption{Component 1 }
    \label{simu1:setting1:comp1}
    \end{subfigure}\hspace{0.5\textwidth}
     \begin{subfigure}{.7\textwidth} 
      \centering
     \includegraphics[width=\textwidth]{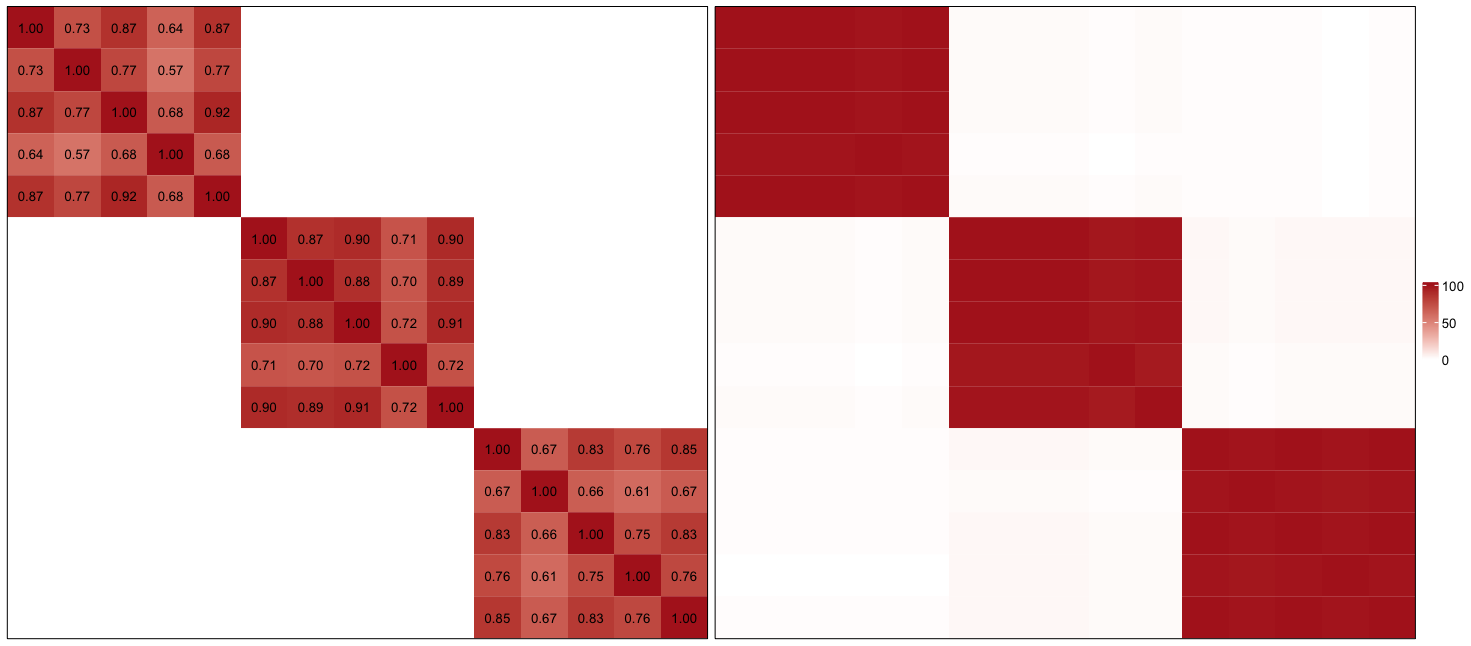}
    \caption{Component 2 }
     \label{simu1:setting1:comp2}
     \end{subfigure}\\
    % \vspace*{-4mm} 
   \caption{True correlation matrix of $\Xv$ (left); estimated correlation matrix of $\Xv$ (right) for each component in Setting 1.}  
     \label{sett1}
\end{figure}

\begin{figure}
\centering
\begin{subfigure}{.7\textwidth}
\centering
\includegraphics[width=\textwidth]{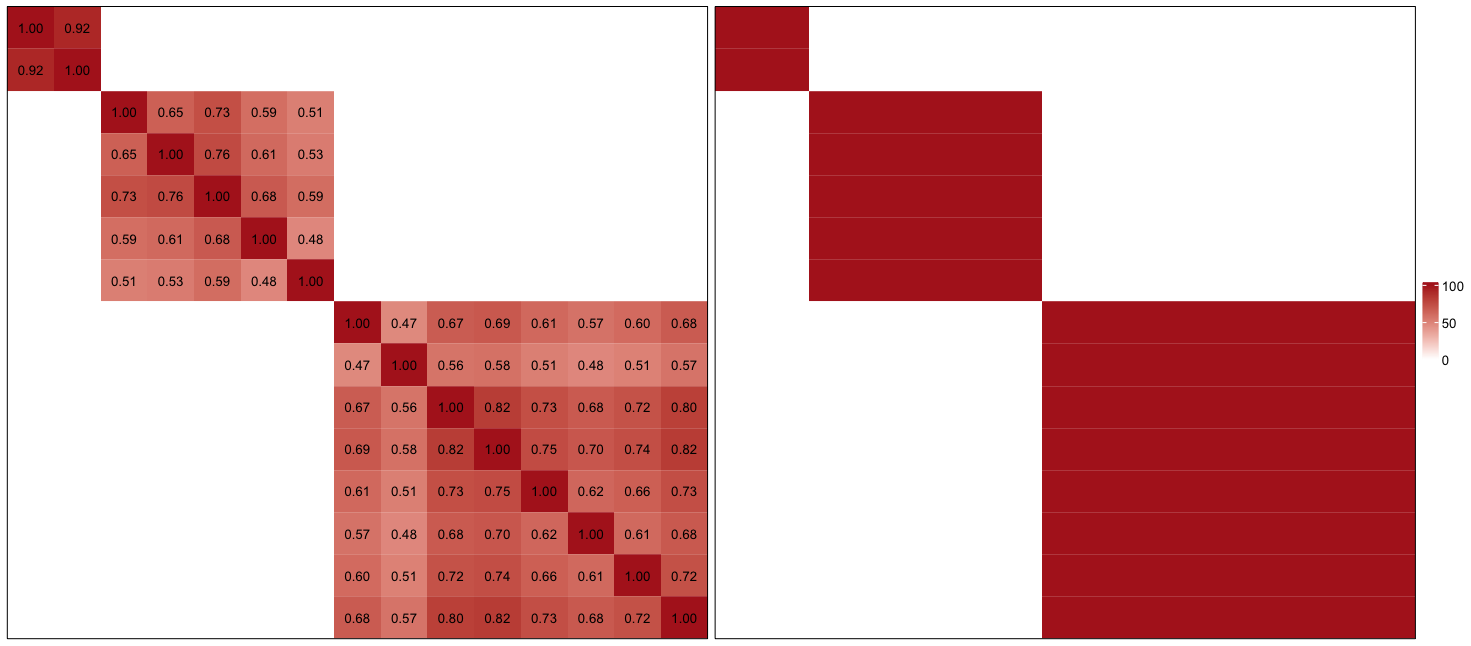}
\caption{Component 1 }
    \label{simu1:setting2:comp1}
    \end{subfigure}\hspace{0.5\textwidth}
     \begin{subfigure}{.7\textwidth} 
      \centering
     \includegraphics[width=\textwidth]{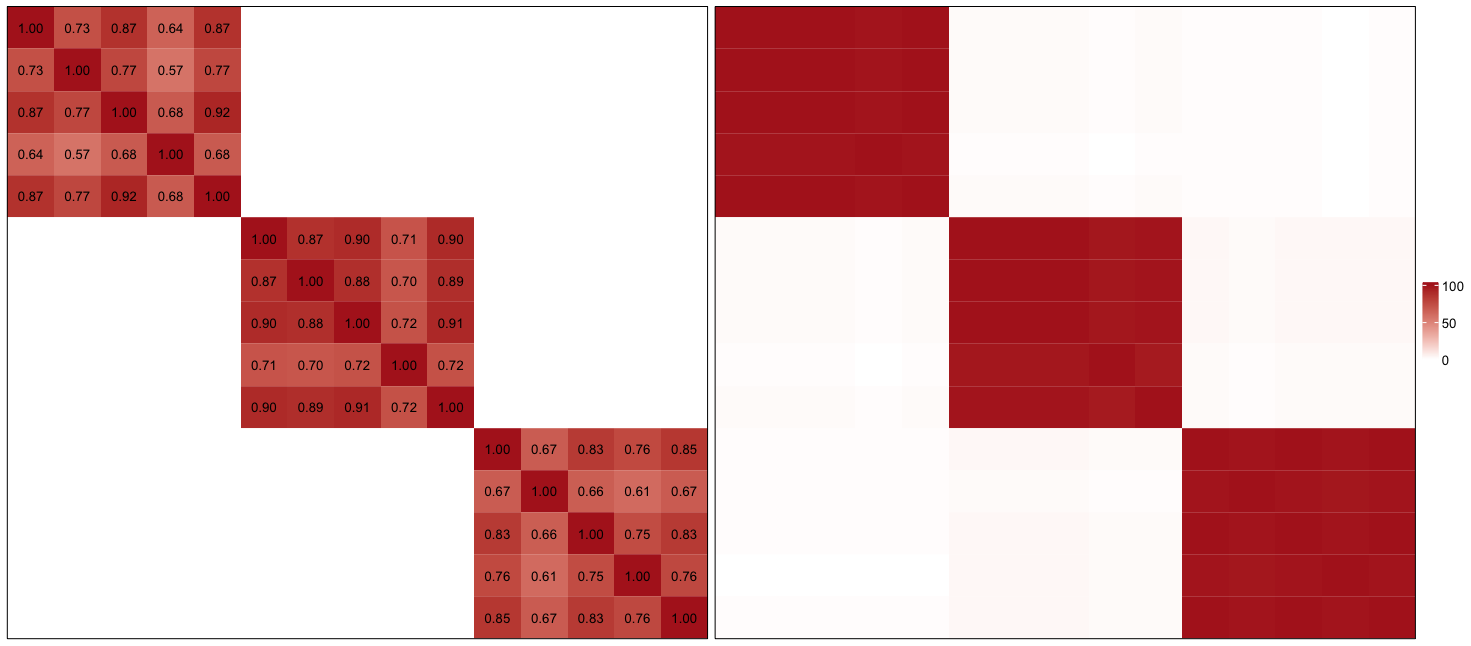}
    \caption{Component 2 }
     \label{simu1:setting2:comp2}
     \end{subfigure}\\
    % \vspace*{-4mm} 
   \caption{True correlation matrix of $\Xv$ (left); estimated correlation matrix of $\Xv$ (right) for each component in Setting 2.}  
     \label{sett2}
\end{figure}

\begin{figure}
\centering
\begin{subfigure}{.7\textwidth}
\centering
\includegraphics[width=\textwidth]{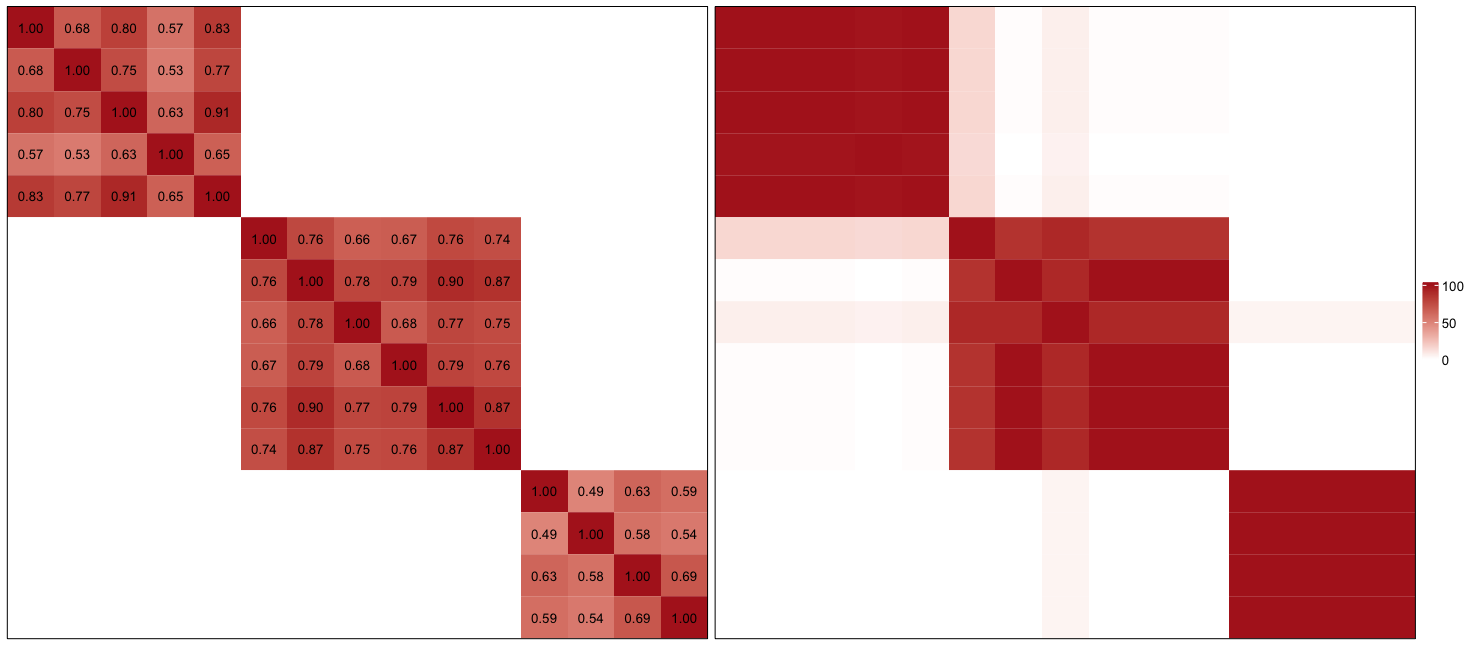}
\caption{Component 1 }
    \label{simu1:setting3:comp1}
    \end{subfigure}\hspace{0.5\textwidth}
     \begin{subfigure}{.7\textwidth} 
      \centering
     \includegraphics[width=\textwidth]{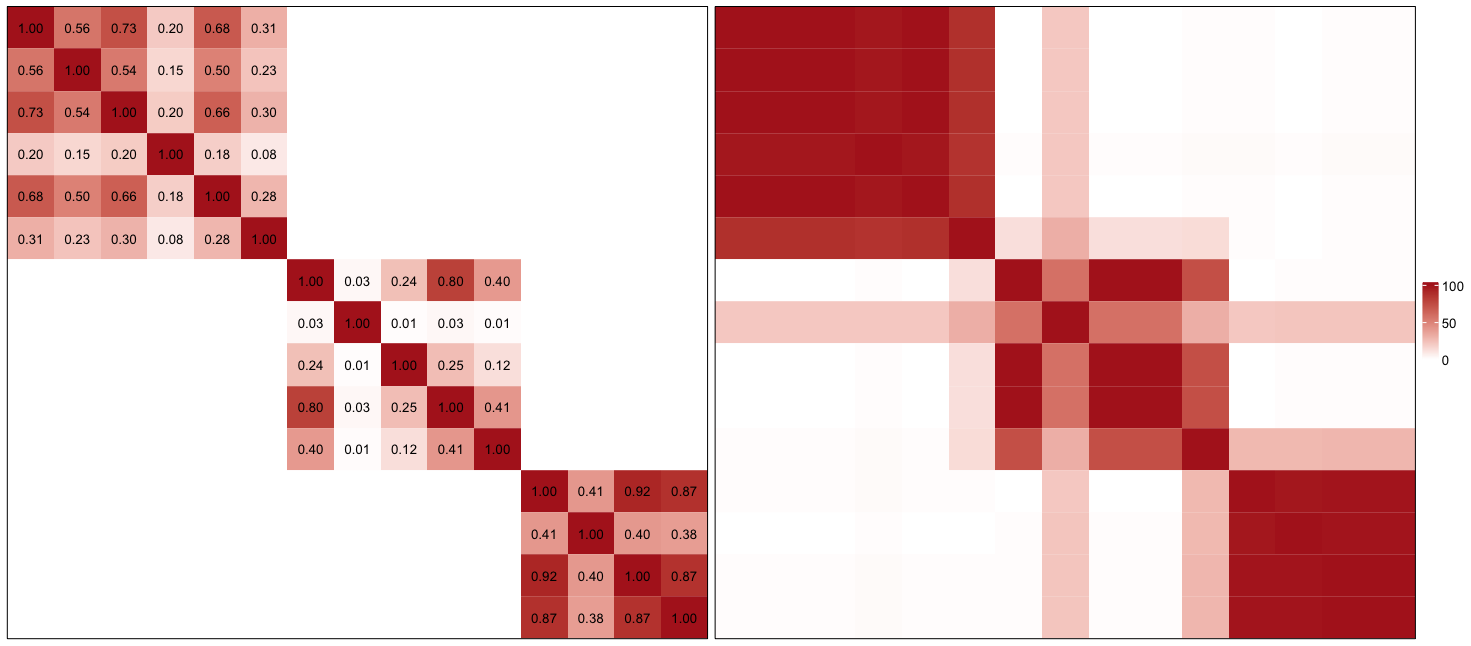}
    \caption{Component 2 }
     \label{simu1:setting3:comp2}
     \end{subfigure}\hspace{0.5\textwidth}
     \begin{subfigure}{.7\textwidth} 
      \centering
     \includegraphics[width=\textwidth]{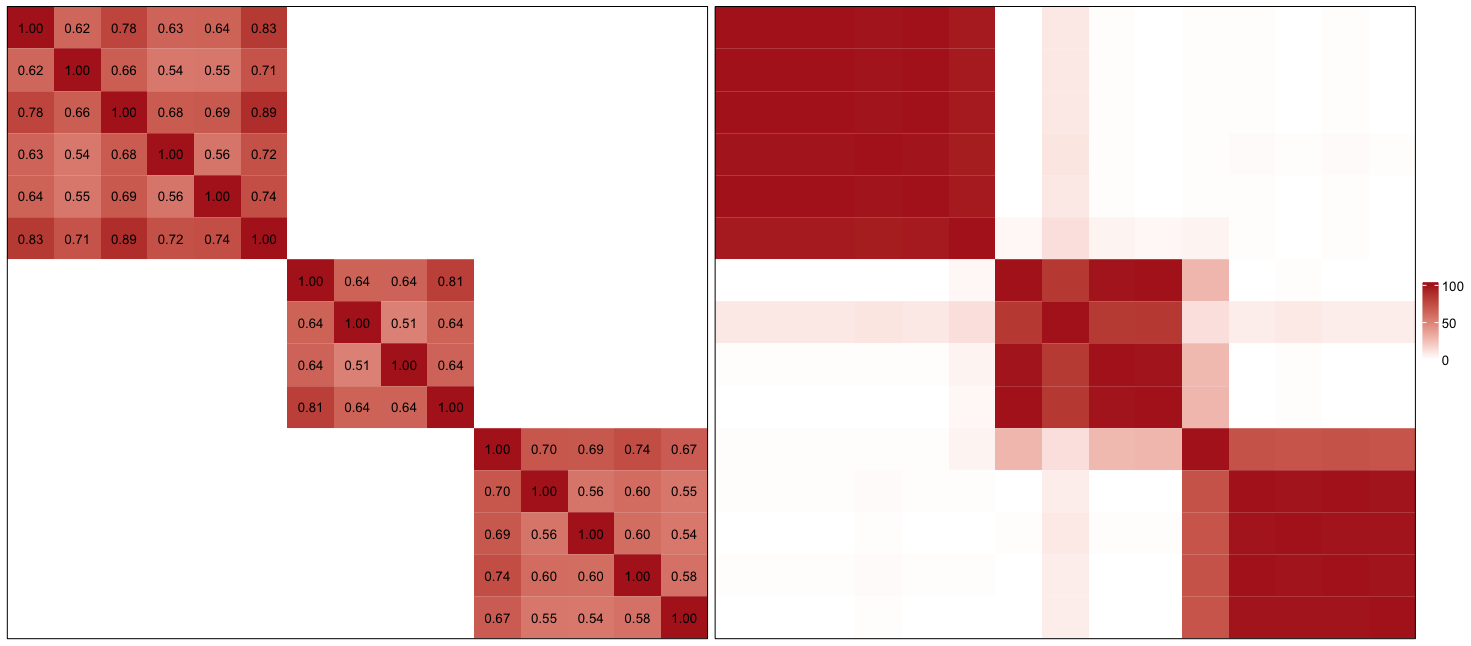}
    \caption{Component 3 }
     \label{simu1:setting3:comp3}
     \end{subfigure}\\
    % \vspace*{-4mm} 
   \caption{True correlation matrix of $\Xv$ (left); recovered groups of explanatory variables out of 100 times (right) for each component in Setting 3.}  
     \label{sett3}
\end{figure}

On the other hand, to investigate the clustering of the explanatory variables, we compared the estimated component-specific correlation structures of $\Xv$ with the true component-specific correlation structures of $\Xv$. The results for Settings 1, 2, and 3 are summarized as heatmaps shown in  Figures  \ref{sett1},  \ref{sett2}, and \ref{sett3}, respectively.\\
As observed in Setting 1 and 2, the true and the estimated correlation matrices are close to each other, indicating a high model's ability to recover the disjoint factors. In Setting 3, while the block structure of correlation matrices for components are close, there are discrepancies as shown by the lighter regions. It is particularly obvious for Component 2.\\
The model appears to struggle more with recovering partitions where the corresponding true correlations are near zero.  This result aligns with our expectations. According to the model's definition, each factor is expected to consist of explanatory variables which tend to be strongly correlated with each other and equally contribute to explaining the response. However, if there are explanatory variables that explain the response as well but are less correlated, the model may struggle to accurately recover the true classification. Also, the smaller sample size of each component in Setting 3 may hamper the accuracy.

Regarding the recovery of component-specific parameters, Tables  \ref{simu1:setting1:MSE},  \ref{simu1:setting2:MSE}, and  \ref{simu1:setting3:MSE} in Appendix \ref{B} display the entrywise average MSEs for Settings 1, 2, and 3, respectively, when the correct model was selected (according to ICL). The results indicate that the deviations from the true parameters are relatively small, suggesting that the model performs well in recovering component-specific parameters.

\subsection{Simulation 2}
In this simulation study, we investigated how varying the number of explanatory variables ($p = 10, 15, 20, 30$) and sample sizes ($N = 500, 750, 1000$) affect the model's performance. Specifically, we simulated 100 datasets considering the block structure as defined in Setting 2 of Simulation 1 and the parameters specified in Table \ref{simu1:setting2:parameters}. Table \ref{tab:varing_np} provides the number of times the true model was selected by AIC, AIC3, BIC and ICL and the average ARI when the true model was selected across 100 datasets for various combinations of  $p$ and $N$.\\
Most of the time the criteria can select the correct models and the average ARIs suggest perfect clustering recovered by the algorithm. BIC and ICL still perform better than AIC or AIC3, consistent with the observation in Simulation 1. The model selection suffers from the high dimensional explanatory variables. Given a fixed sample size, the criteria are more likely to identify the true components when the dimension is low.    
 
%\textbf{Table  \ref{tab:varing_np} needs to be commented.}
\begin{table}[H]
    \centering
    \begin{tabular}{c|c|ccccc}
                                  &         & AIC(ARI)  & AIC3(ARI) & BIC(ARI)  & ICL(ARI) \\ \hline
         \multirow{3}{*}{$N$ = 500} &  $p$ = 10 & 98(1.00) & 100(1.00) & 100(1.00) & 100(1.00) \\
                                    &  $p$ = 15 & 99(1.00) & 100(1.00) & 100(1.00) & 100(1.00) \\
                                    &  $p$ = 20 & 91(1.00) & 95(1.00)  & 97(1.00)  & 97(1.00)\\
                                    &  $p$ = 30 & 72(1.00) & 82(1.00)  & 86(1.00)  & 86(1.00)\\ \hline
         \multirow{3}{*}{$N$ = 750} &  $p$ = 10 & 97(1.00) & 100(1.00) & 100(1.00) & 100(1.00)\\
                                    &  $p$ = 15 & 93(1.00) & 100(1.00) & 100(1.00) & 100(1.00)\\
                                    &  $p$ = 20 & 95(1.00) & 100(1.00) & 100(1.00) & 100(1.00)\\
                                    &  $p$ = 30 & 72(1.00) &  87(1.00) &  90(1.00) &  90(1.00)\\ \hline
         \multirow{3}{*}{$N$ = 1000}&  $p$ = 10 & 88(1.00) & 100(1.00) & 100(1.00) & 100(1.00)\\
                                    &  $p$ = 15 & 84(1.00) & 100(1.00) & 100(1.00) & 100(1.00)\\
                                    &  $p$ = 20 & 89(1.00) &  97(1.00) &  98(1.00) &  98(1.00)\\
                                    &  $p$ = 30 & 67(1.00) &  87(1.00) &  90(1.00) &  90(1.00)\\     \end{tabular}
    \caption{The number of times the correct model was selected of model criteria out of 100 datasets and ARI values (in brackets).}
    \label{tab:varing_np}
\end{table}

\subsection{Simulation 3}
Since in real data, the covariance/correlation matrix cannot be a perfect block diagonal matrix due to natural noise and overlap between explanatory variables, in this final simulation study, we aim to investigate how relatively small correlations between explanatory variables in different clusters (i.e. non-zero off-diagonal entries) influence the model's performance.
Here we set the off-diagonal entries to non-zero but without exceeding the correlations in diagonal blocks. To control the level of noise under $m$ while preserving the semi-positive definite correlation structure, we generate the covariance matrix $\Psiv_g$ in the following way:
\begin{enumerate}
   \item Generate a symmetric matrix $\Phiv_g$ whose entry $(\Phiv_g)_{ij} = U_{ij}\sqrt{(\Gammav_g)_{ii}(\Gammav_g)_{jj}}$, where $(\Gammav_g)_{ii}$, $(\Gammav_g)_{jj}$ are entries of $\Gammav_g$, and $U_{ij}$ is drawn from a uniform distribution $\mathrm{U}(-m, m)$. We set $U_{ij} = U_{ji}$ for symmetry. Here $m>0$ is used as the level of noise.
   \item Do eigendecomposition of $\Phiv_g = \Pv\Ev\Pv'$ as $\Phiv$ is symmetric. Here $\Pv$ is the matrix of eigenvectors and $\Ev$ is a diagonal matrix corresponding to eigenvalues. Also, we have $\Pv' = \Pv^{-1}$.
   \item Reconstruct $\widetilde{\Phiv}_{g} = \Pv\widetilde{\Ev}\Pv'$ where the entry \begin{equation*}
       \widetilde{\Ev}_{ij} = 
       \begin{cases}
        \Ev_{ij}, \Ev_{ij} \geq 0\\
        0, \textrm{otherwise.}
        \end{cases}
   \end{equation*}
   \item Create $\Psiv_g$ whose entry $(\Psiv_g)_{ij}$ is 
   \begin{equation*}
       (\Psiv_g)_{ij} = \begin{cases}
           1 + (\widetilde{\Phiv}_g)_{ii}, i = j\\
           (\widetilde{\Phiv}_g)_{ij}, \textrm{otherwise,}\\
       \end{cases}
   \end{equation*}
   which guarantees a positive definite correlation.
\end{enumerate}

\noindent To see the effect of noise level $m$, we deploy two settings. In Setting 1, $m$ is $0.43$; in Setting 2, $m$ is $0.77$. Then we simulated 100 datasets with $G = 2$ components, $M = 5$ response variables, $p = 20$ explanatory variables, $Q = 3$ segments of explanatory variables and sample size $n = 750$. The mixing proportion is $\pi = (0.6, 0.4)$.  We ran the AECM algorithm on each dataset with $G = 1, 2, 3, 4, 5$ and $Q = 1, 2, 3, 4, 5$, respectively. 

\begin{figure}[H]
    \centering
    \begin{minipage}{\linewidth}
    \includegraphics[width=\linewidth]{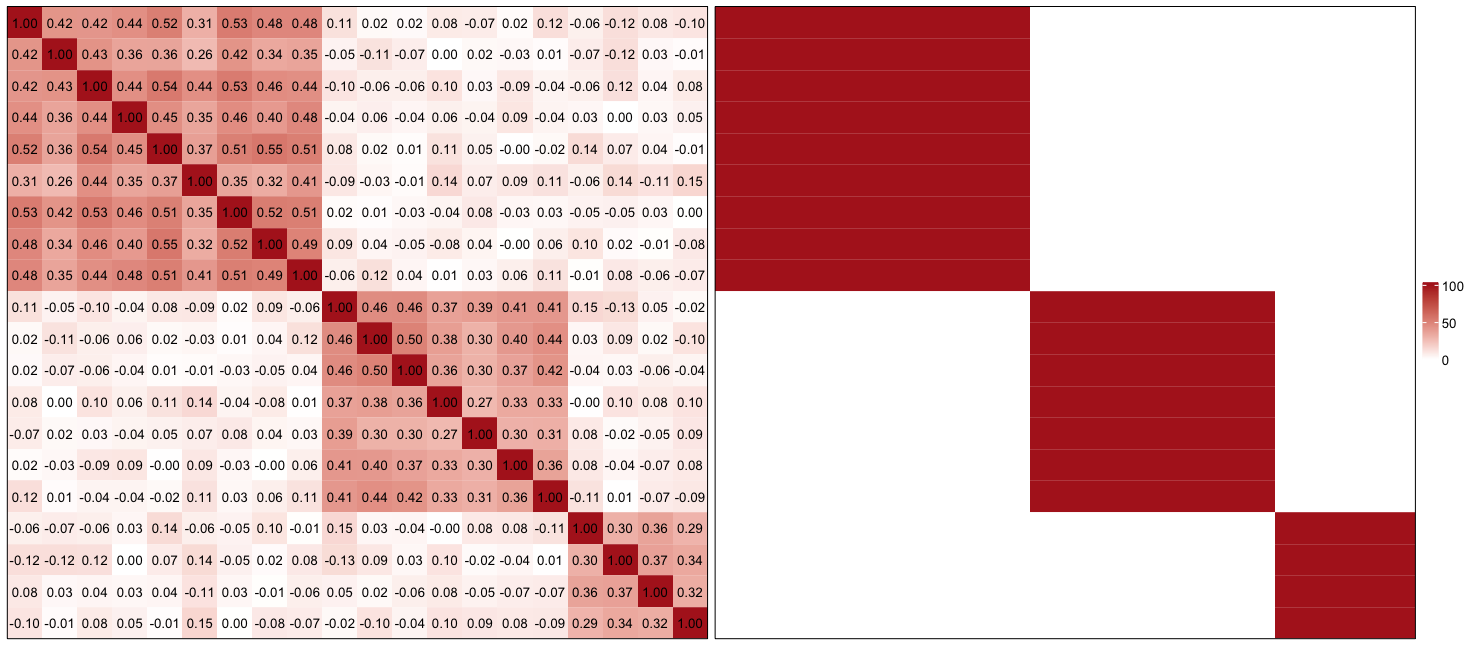}
    \captionof*{figure}{(A) Component 1}
    \end{minipage}\\
    \begin{minipage}{\linewidth}
    \includegraphics[width=\linewidth]{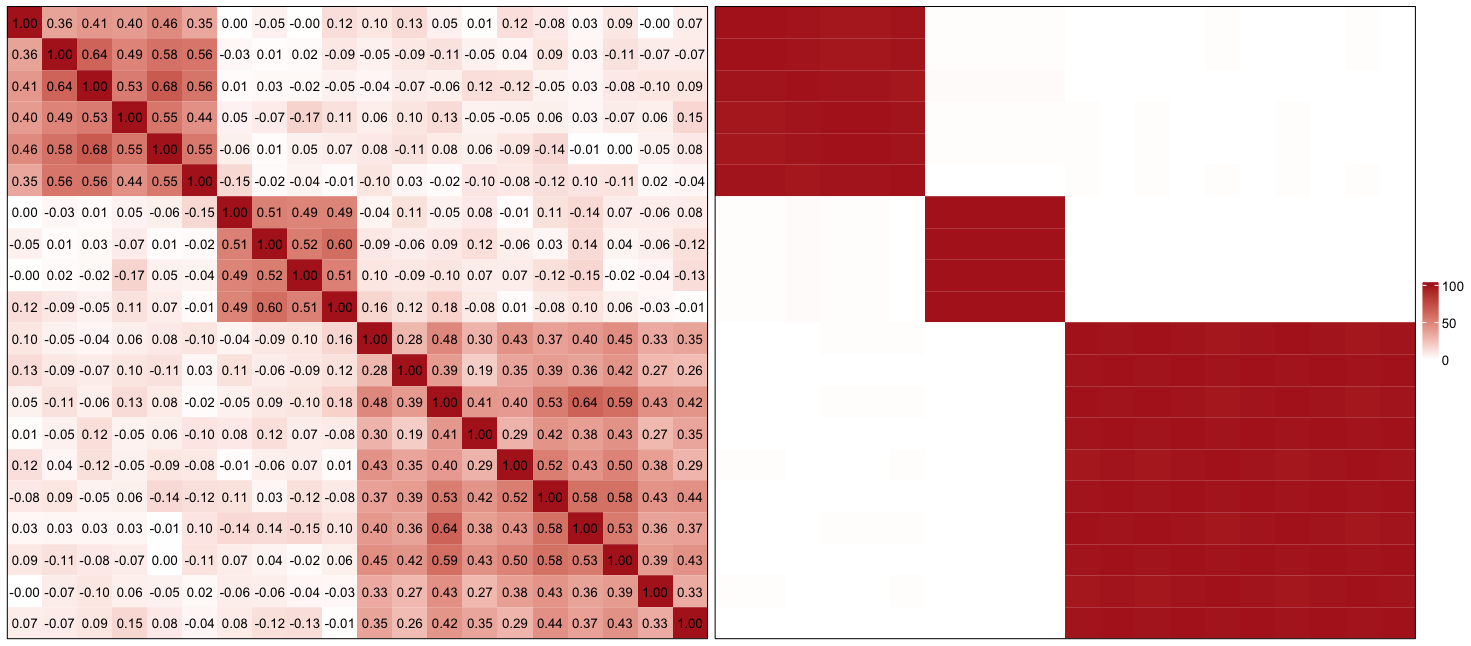}
    \captionof*{figure}{(B) Component 2}
    \end{minipage}%
    \caption{True correlation matrix of $\Xv$ (left); recovered groups of explanatory variables out of 100 times (right) for each component.}
    \label{simu3:setting1}
\end{figure}

\begin{figure}[H]
    \centering
    \begin{minipage}{\linewidth}
    \includegraphics[width=\linewidth]{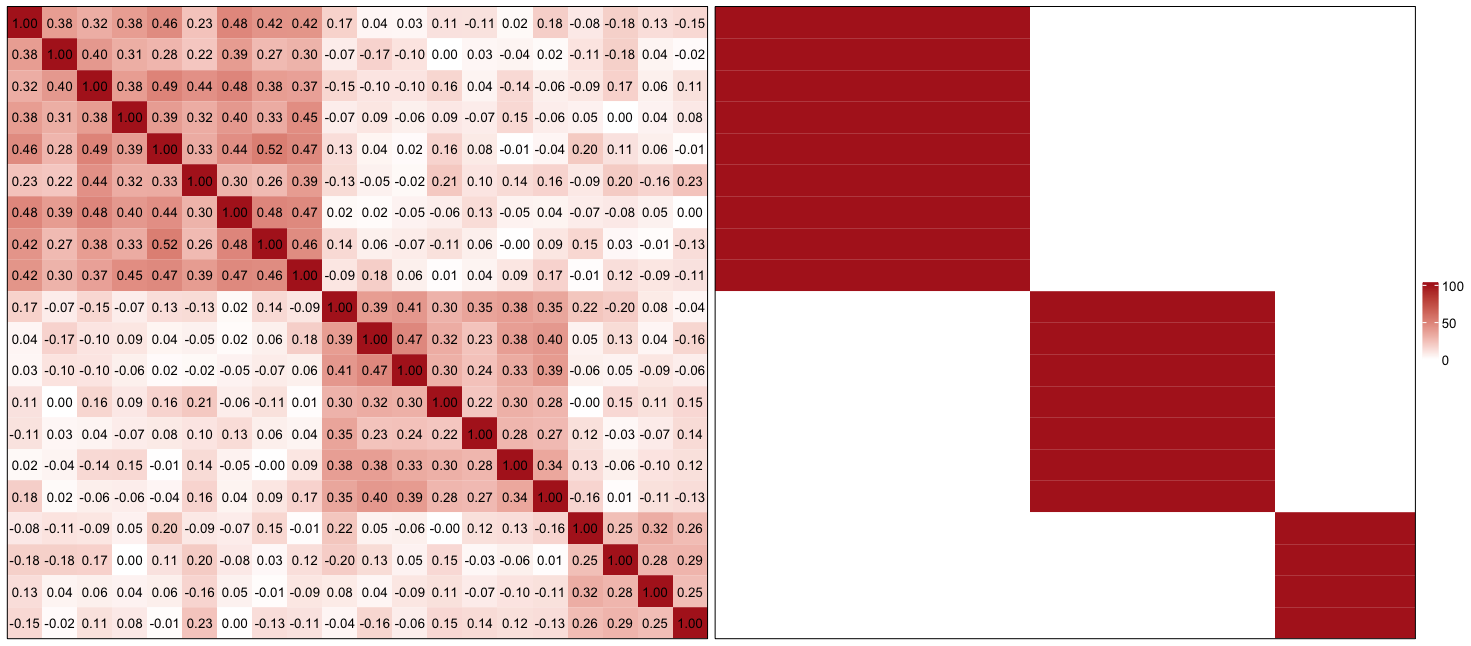}
    \captionof*{figure}{(A) Component 1}
    \end{minipage}\\
    \begin{minipage}{\linewidth}
    \includegraphics[width=\linewidth]{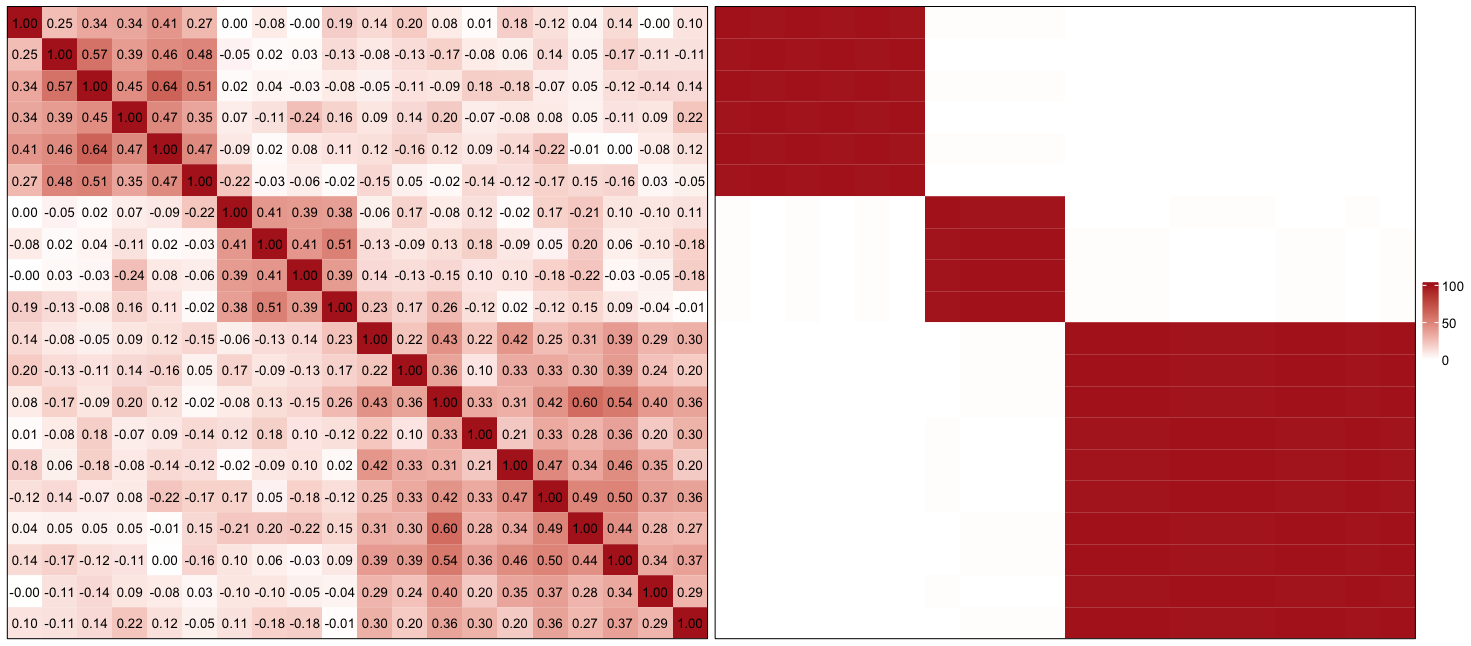}
    \captionof*{figure}{(B) Component 2}
    \end{minipage}%
    \caption{True correlation matrix of $\Xv$ (left); recovered groups of explanatory variables out of 100 times (right) for each component.}
    \label{simu3:setting2}
\end{figure}

The recovered segments/groups of explanatory variables are obtained for each dataset and summarised in Figure \ref{simu3:setting1} and \ref{simu3:setting2}. The clustering of explanatory variables accurately captures the main structure of the true correlation matrix in both settings. 

\section{Crime data of communities within the United States}
\label{sec:7}
In this section, we start by performing an exploratory analysis of the real-world dataset named `Community and Crime Unnormalized' from the UCI Machine Learning Repository. Subsequently, we apply the proposed model to the dataset and discuss the results obtained.

\subsection{Exploratory data analysis}
The crime dataset combines socioeconomic information of communities from the 1990 United States Census, law enforcement data from the 1990 Law Enforcement Management and Admin Stats survey, and crime data from the 1995 FBI UCR.  
Each dataset entry is an observation at the community level, including rates of different types of crime, population demographics, age structure, socio-economic status, etc. 
In this study, we focused exclusively on economic crimes ($M=3)$, including robberies, burglaries, and larcenies, among all types of offences. Additionally, we examined various community-level characteristics related to (a) general population demographics, (b) income and employment, and (c) housing.  After excluding highly collinear socio-economic features, we considered the remaining $p=16$ socio-economic variables, encompassing  $N$= 1901 communities (observations with missing records were excluded from the analysis).  A brief overview of the variables included in the analysis is provided in Table  \ref{tab:variables}. Skewed variables were log-transformed to meet the normality assumptions.
\begin{table}[H]
    \tiny
    \centering
    \begin{tabular}{c|c|c|c}
    \toprule
         \textbf{Type} & \textbf{General group} & \textbf{Varaible Name}& \textbf{Description}\\
         \hline
         \multirowcell{3}{Responses \\ Variables} & \multirowcell{3}{Properties related\\ crimes} & Robberies&Number of robberies per 100,000 population\\ \cline{3-4}
         &&Burglaries& Number of burglaries per 100,000 population\\ \cline{3-4}
         &&Larcenies & Number of larcenies per 100,000 population\\ \cline{3-4}
         \hline
         \multirowcell{16}{Explanatory \\ variables} & \multirowcell{2}{General\\ Characteristics}  & pop &\\ \cline{3-4}
         && pct\_low\_edu & \makecell{Percentage of people 25 and over with \\less than a 9th grade education}\\\cline{2-4}
         &\multirowcell{6}{Income \\and\\ Employment}& pct\_retire &Percentage with retirement income\\\cline{3-4}
         && pct\_wsocsec & Percentage with SSI\\\cline{3-4}
         && pct\_unemploy & Percentage of unemployment\\\cline{3-4}
         && pct\_wfarm & \makecell{Percentage of households with \\farm or self employment income}\\\cline{3-4}
         && pct\_employ\_mfg &\makecell{Percentage of people \\in manufacturing}\\ \cline{3-4}
         &&pct\_employ\_prof\_serv &\makecell{Percentage of people \\in professional services}\\ \cline{2-4}
         &\multirow{7}{*}{Housing related}&pct\_vacant\_boarded & \makecell{Percentage of vacant housing boarded up}\\\cline{3-4}
         && pct\_vacant6up &\makecell{Percentage of housing that\\ has been vacant more than 6 months}\\\cline{3-4}
         && pct\_hous\_owner\_occup & \makecell{Percentage of households owner occupied}\\\cline{3-4}
         && own\_hous\_qrange &Q range of owner-occupied housing \\\cline{3-4}
         && rent\_qrange & Q range of rent housing\\\cline{3-4}
         && med\_rentpct\_hous &\makecell{Median rent as a percentage\\ of household income}\\\cline{3-4}
         && med\_own\_costpct & \makecell{Median owners cost as a percentage of \\household income for owners}\\\cline{3-4}
         && med\_own\_cost\_pct\_wo & \makecell{Median owners cost as a percentage of \\household income \\for owners without a mortgage }\\\bottomrule
    \end{tabular}
    \caption{Details of the variables included in the analysis.}
    \label{tab:variables}
\end{table}

\begin{figure}[!h]
    \centering
    \includegraphics[width = \linewidth]{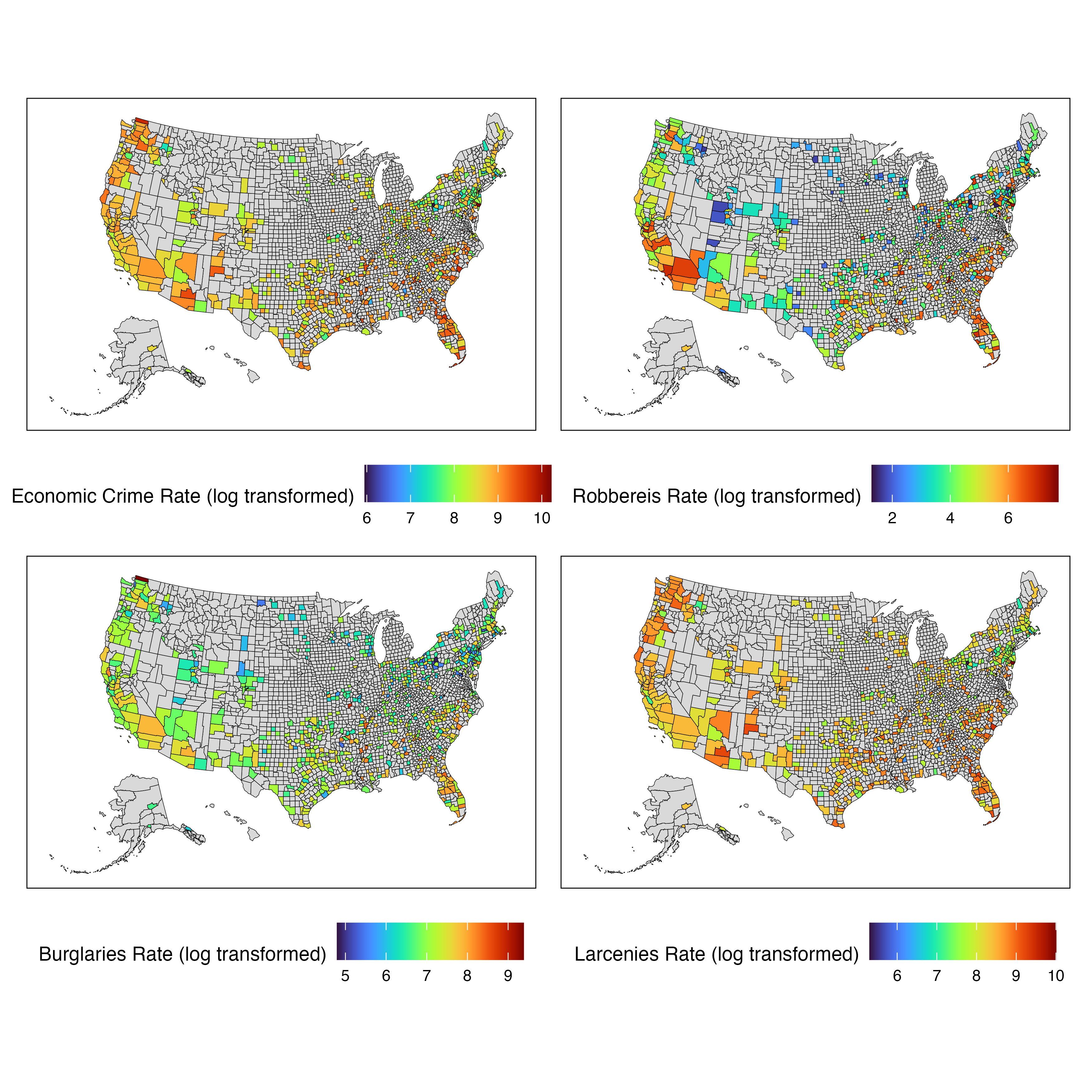}
    \caption{The distribution of log-transformed economic crime rates per 100k population at the county level.}
    \label{crime_distribution}
\end{figure}

For a more comprehensive understanding of economic crimes, Figure \ref{crime_distribution} presents the distributions of log-transformed crime rates at the county level.  As observed, the analysis of these distributions confirms significant heterogeneity in economic crimes across the US. 
The distribution of robberies has a quite clear pattern: we can observe the relatively high values on the West Coast and the Southeastern region, which hints at a decreasing trend from the coasts to the central and from the south to the north on the East Coast.
Similarly, burglary rates follow a similar pattern, with higher numbers in California and Florida compared to a more uniform distribution across the rest of the US.
Furthermore, the number of burglaries is much higher in Whatcom County of Washington compared to its adjacent region.
The larcenies are the most frequent economic crimes of the three crimes we investigated. Most counties have a high rate of larceny, however, several exceptions can be found in the Northeast and the South. Larceny dominates the overall economic crime rate because of its high frequency. So, the pattern of overall crime rates is similar to larcenies.

\subsection{Results from the MCWDFA Model}
Driven by the well-known diverse crime landscape across communities in the United States, we aim to investigate how the $p=16$ socio-economic factors impact the $M=3$ crime rates. Specifically, our goal is twofold: first, to identify community groups with distinct effects of socio-economic factors on crime rates, and second, to determine if there are segments or groups of socio-economic factors that similarly influence economic crimes.
We applied the proposed AECM algorithm with $G = 1, 2, \dots, 10$ and $Q = 1, 2, \dots, 10$ for ten times. The model with the highest ICL was selected ($G=6, Q=3$).

\begin{figure}
\vspace{-0.1in}
   \centering
      \begin{subfigure}{\textwidth}
        \centering
   \includegraphics[width =0.65\linewidth]{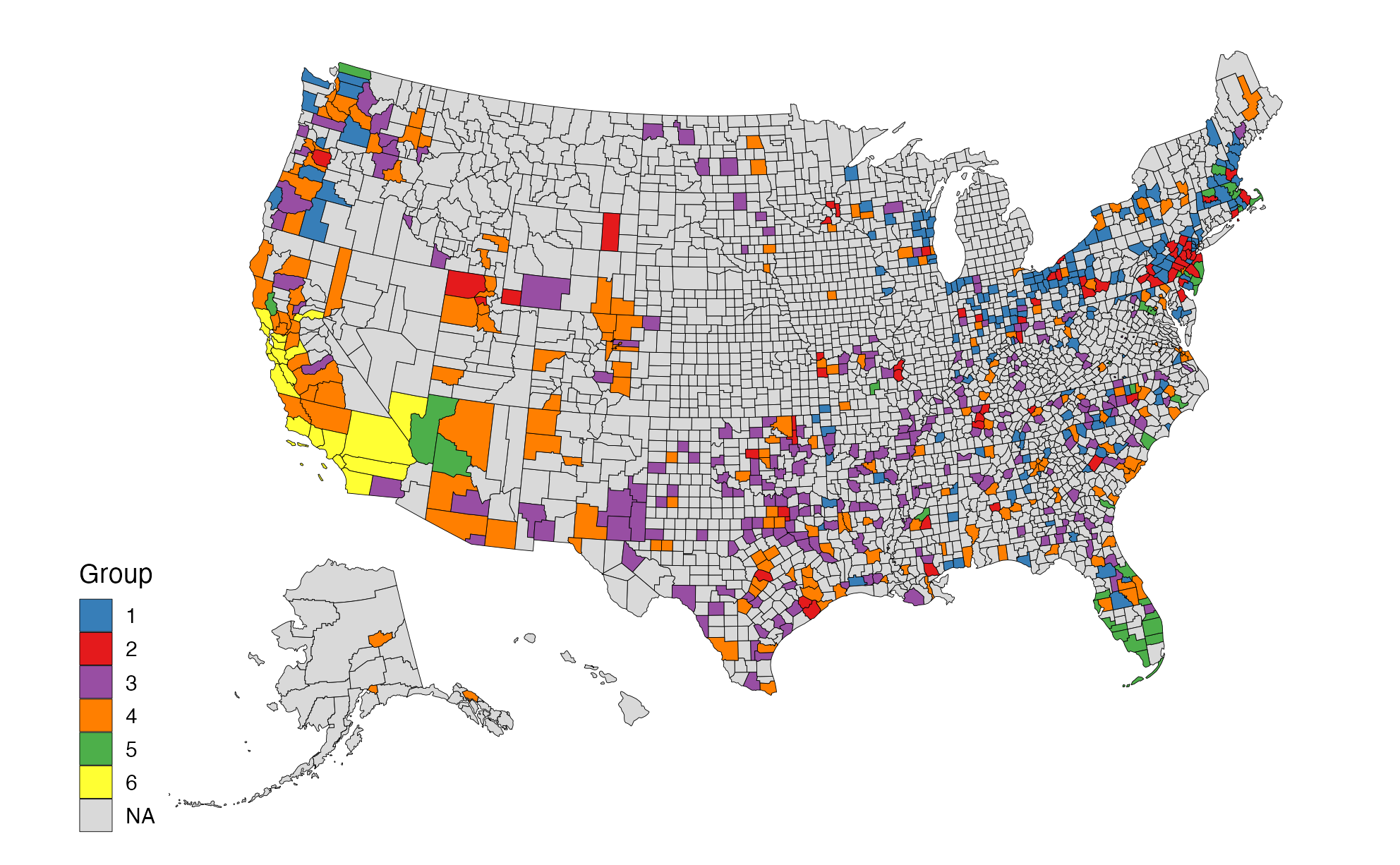}
    \caption{Community clusters visualized at the county level on the US map}
    \label{usmap}
    \end{subfigure}\\
    \begin{subfigure}{0.5\textwidth}
        \centering
        \includegraphics[width=0.75\linewidth]{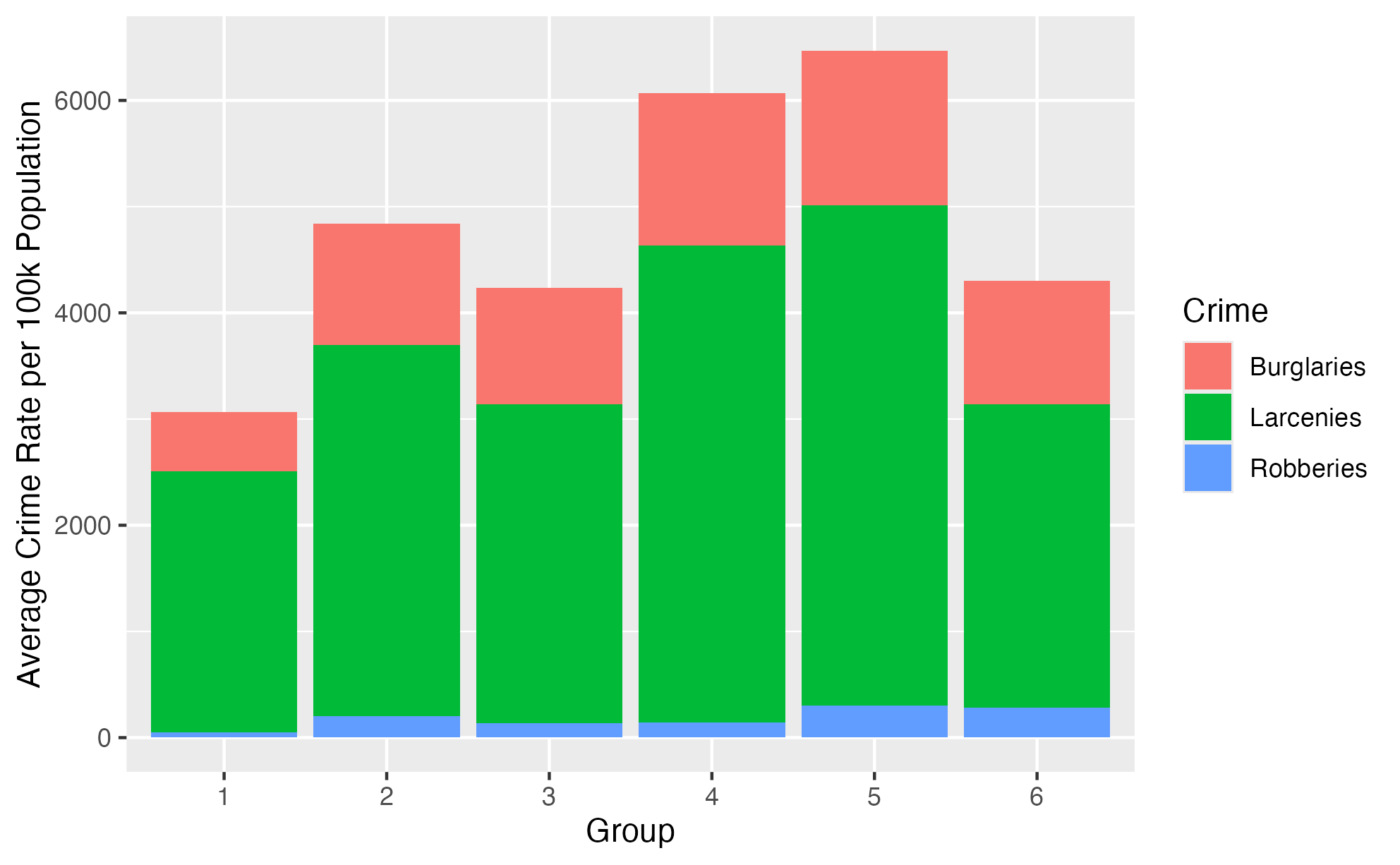}
        \caption{Economic crime rates by clusters}
        \label{crime_rates}
    \end{subfigure}%
    \\
    \begin{subfigure}{\textwidth}
        \centering
        \includegraphics[width=0.7\linewidth]{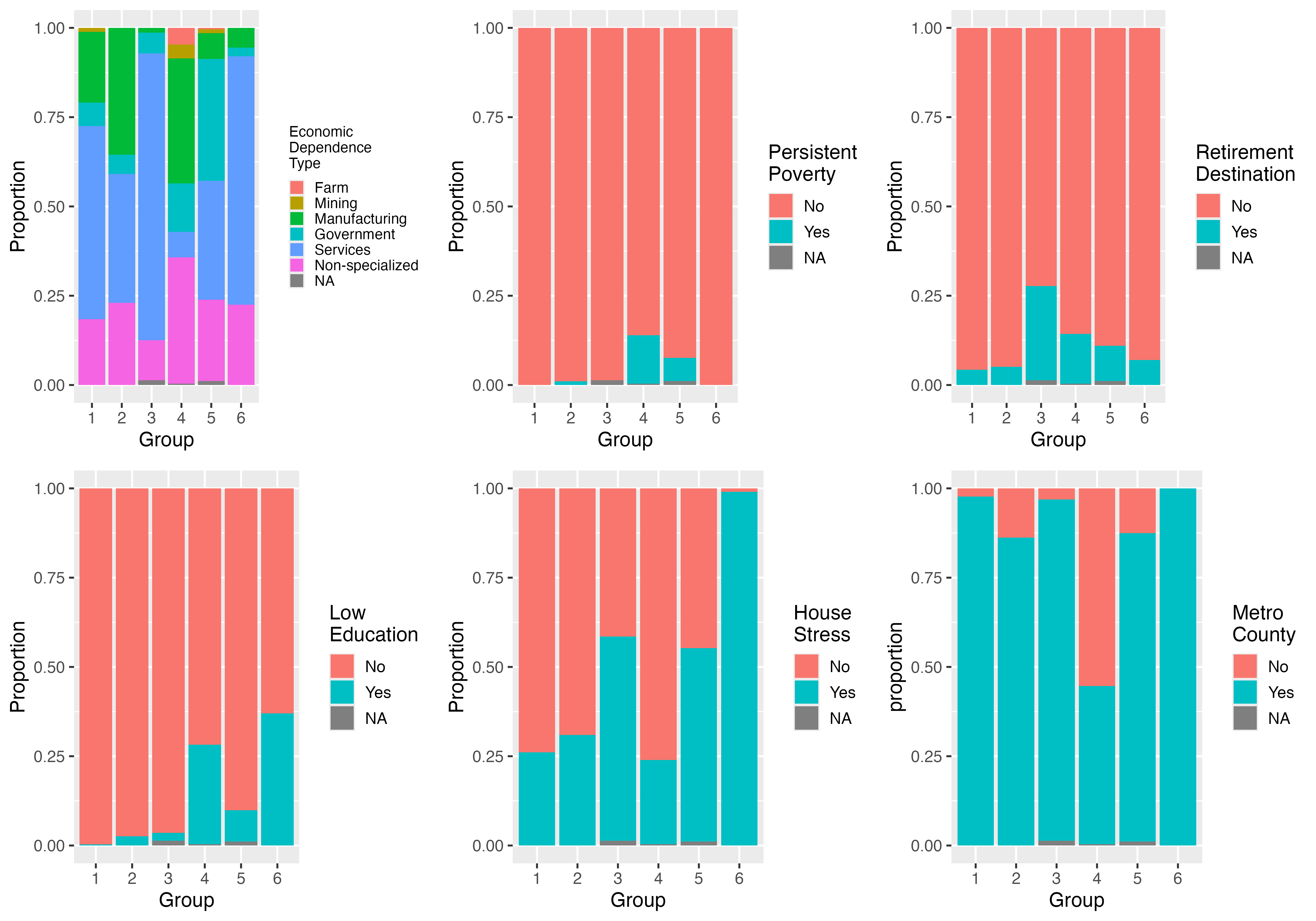}
        \caption{Characteristics of the clusters}
        \label{typology}
    \end{subfigure}
    \caption{Clustering results at the county level: (a) visualization of clusters on the US map, (b) economic crime rates by clusters, and (c) characteristics of the clusters.}
\end{figure}

Figure \ref{usmap} provides a visualization of the clustering results at the county level on the US map. The average crime rates of the estimated clusters are presented in Figure \ref{crime_rates}. To characterize the estimated clusters, we also use external county classification data provided by the Economic Research Service U.S. Department of Agriculture Economic Research Services (https://www.ers.usda.gov/data-products/county-typology-codes/)  to better understand the socio-economic and geographic contexts of each cluster.
We considered the following external variables: the economy types the counties depend on, whether the county is a retirement destination, the presence of persistent poverty, levels of education, employment rates, and significant housing stress. The characterization of the clusters based on these variables is summarized in Figure \ref{typology}.

The main findings are summarized as follows. Counties in Cluster 1 and Cluster 2 are concentrated in the Northeastern. They differ mainly because of varying levels of crime: Cluster 2 has higher crime rates compared to Cluster 1, and Cluster 1 has the lowest overall crime rates. Cluster 2 has a higher proportion of manufacturing-dependent counties and lower services-dependent counties.
Cluster 3, on the other hand, has the most retirement destinations, the highest proportion of services-dependent
counties, fewer counties with persistent child poverty and they are almost all in metro counties. Cluster 4 and Cluster 5 both have high crime rates but differ significantly. Cluster 4 has the highest proportion of non-metro counties (likely rural), with persistent child poverty, low education, and a high proportion of manufacturing, farm and mining-dependent counties. In contrast, Cluster 5 has a high proportion of government-dependent counties, more counties with housing stress, and fewer counties with low education.
Cluster 6 predominantly includes metropolitan counties in the western United States, particularly California and Nevada. It is characterized by average crime rates and the highest level of counties believed low education. This cluster primarily consists of service-dependent counties that face significant housing stress and does not include any counties experiencing persistent poverty.
In summary, our results are consistent with established geographic and economic patterns, indicating that the composition of the clusters reflects meaningful geographic and economic distinctions.\\

\begin{table}[ht]
\centering\tiny
% \arrayrulecolor{black} % Set the color of the table lines to black
\resizebox{\textwidth}{!}{%
\begin{tabular}{l|c|l|c|l|c}
\hline
% \rowcolor{gray!30}
\multicolumn{2}{|c|}{\textbf{Cluster 1}} & \multicolumn{2}{c|}{\textbf{Cluster 2}} & \multicolumn{2}{c|}{\textbf{Cluster 3}} \\
\hline
\textbf{Factor 1:}&\textbf{Weight}&\textbf{Factor 1: }&\textbf{Weight}&\textbf{Factor 1: }&\textbf{ Weight} \\
  {\color{red}{pct\_wsocsec}} & 0.43 & {\color{red}{pct\_wsocsec}} & 0.83 & {\color{patriarch}{pop}} & 0.01 \\ 
  {\color{red}{pct\_retire}} & 0.76 & {\color{red}{pct\_retire}} & 0.86 & {\color{patriarch}{pct\_low\_edu}} & 0.36 \\ 
  {\color{red}{pct\_employ\_mfg}} & 0.03 & {\color{patriarch}{pct\_low\_edu}} & 0.41 & {\color{red}{pct\_unemploy}} & 0.77 \\ 
  {\color{brightgreen}{pct\_hous\_owner\_occup}} & 0.43 & {\color{red}{pct\_unemploy}} & 0.16 & {\color{brightgreen}{pct\_vacant\_boarded}} & 0.35 \\ 
  \textbf{Factor 2:} &  & {\color{red}{pct\_employ\_prof\_serv}} & 0.19 & {\color{brightgreen}{rent\_qrange}} & 0.10 \\ 
  {\color{red}{pct\_wfarm}} & 0.01 & \textbf{Factor 2:} &  & {\color{brightgreen}{med\_rentpct\_hous}} & 0.59 \\ 
  {\color{brightgreen}{own\_hous\_qrange}} & 0.58 & {\color{patriarch}{pop}} & 0.01 & {\color{brightgreen}{med\_own\_costpct}} & 0.40 \\ 
  {\color{brightgreen}{rent\_qrange}} & 0.76 & {\color{red}{pct\_wfarm}} & 0.10 & {\color{brightgreen}{med\_own\_cost\_pct\_wo}} & 0.21 \\ 
  {\color{brightgreen}{med\_own\_costpct}} & 0.90 & {\color{brightgreen}{pct\_hous\_owner\_occup}} & 0.36 & \textbf{Factor 2:} &  \\ 
  {\color{brightgreen}{med\_own\_cost\_pct\_wo}} & 0.71 & {\color{brightgreen}{pct\_vacant\_boarded}} & 0.20 & {\color{red}{pct\_wfarm}} & 0.19 \\ 
  \textbf{Factor 3:} &  & {\color{brightgreen}{pct\_vacant6up}} & 0.40 & {\color{brightgreen}{pct\_hous\_owner\_occup}} & 0.42 \\ 
  {\color{patriarch}{pop}} & 0.23 & {\color{brightgreen}{own\_hous\_qrange}} & 0.69 & {\color{brightgreen}{pct\_vacant6up}} & 0.38 \\ 
  {\color{patriarch}{pct\_low\_edu}} & 0.20 & {\color{brightgreen}{rent\_qrange}} & 1.03 & \textbf{Factor 3:} &  \\ 
  {\color{red}{pct\_unemploy}} & 0.67 & {\color{brightgreen}{med\_rentpct\_hous}} & 0.44 & {\color{red}{pct\_wsocsec}} & 0.44 \\ 
  {\color{red}{pct\_employ\_prof\_serv}} & 0.18 & {\color{brightgreen}{med\_own\_costpct}} & 0.60 & {\color{red}{pct\_retire}} & 0.59 \\ 
  {\color{brightgreen}{pct\_vacant\_boarded}} & 0.61 & {\color{brightgreen}{med\_own\_cost\_pct\_wo}} & 0.53 & {\color{red}{pct\_employ\_mfg}} & 0.02 \\ 
  {\color{brightgreen}{pct\_vacant6up}} & 0.40 & \textbf{Factor 3:} &  & {\color{red}{pct\_employ\_prof\_serv}} & 0.30 \\ 
  {\color{brightgreen}{med\_rentpct\_hous}} & 0.44 & {\color{red}{pct\_employ\_mfg}} & 0.47 & {\color{brightgreen}{own\_hous\_qrange}} & 0.05 \\

\hline
\hline
\hline
\multicolumn{2}{|c|}{\textbf{Cluster 4}} & \multicolumn{2}{c|}{\textbf{Cluster 5}} & \multicolumn{2}{c|}{\textbf{Cluster 6}} \\
\hline                                                       
    \textbf{Factor 1:}             &      \textbf{Weight}    &     \textbf{Factor 1:}             &     \textbf{Weight }    &       \textbf{Factor 1:}            &      \textbf{Weight}         \\
  {\color{red}{pct\_wfarm}} & 0.27 & {\color{red}{pct\_wsocsec}} & 1.39 & {\color{patriarch}{pop}} & 0.18 \\ 
  {\color{red}{pct\_employ\_prof\_serv}} & 1.26 & {\color{red}{pct\_retire}} & 1.18 & {\color{patriarch}{pct\_low\_edu}} & 1.17 \\ 
  {\color{brightgreen}{med\_rentpct\_hous}} & 0.74 & {\color{patriarch}{pct\_low\_edu}} & 0.36 & {\color{red}{pct\_unemploy}} & 0.98 \\ 
  \textbf{Factor 2:} &  & {\color{red}{pct\_unemploy}} & 0.35 & {\color{red}{pct\_employ\_mfg}} & 0.35 \\ 
  {\color{brightgreen}{own\_hous\_qrange}} & 0.54 & {\color{brightgreen}{pct\_hous\_owner\_occup}} & 0.33 & {\color{brightgreen}{pct\_vacant\_boarded}} & 0.47 \\ 
  {\color{brightgreen}{rent\_qrange}} & 0.37 & {\color{brightgreen}{med\_rentpct\_hous}} & 0.55 & {\color{brightgreen}{pct\_vacant6up}} & 0.12 \\ 
  {\color{brightgreen}{med\_own\_costpct}} & 0.24 & \textbf{Factor 2:} &  & {\color{brightgreen}{med\_rentpct\_hous}} & 0.50 \\ 
  \textbf{Factor 3:} &  & {\color{red}{pct\_employ\_mfg}} & 0.10 & {\color{brightgreen}{med\_own\_costpct}} & 0.14 \\ 
  {\color{patriarch}{pop}} & 0.55 & {\color{brightgreen}{pct\_vacant\_boarded}} & 0.17 & {\color{brightgreen}{med\_own\_cost\_pct\_wo}} & 0.07 \\ 
  {\color{red}{pct\_wsocsec}} & 0.46 & {\color{brightgreen}{pct\_vacant6up}} & 0.40 & \textbf{Factor 2:} &  \\ 
  {\color{red}{pct\_retire}} & 0.17 & {\color{brightgreen}{med\_own\_costpct}} & 0.14 & {\color{red}{pct\_wsocsec}} & 0.46 \\ 
  {\color{patriarch}{pct\_low\_edu}} & 0.50 & {\color{brightgreen}{med\_own\_cost\_pct\_wo}} & 1.15 & {\color{red}{pct\_retire}} & 0.71 \\ 
  {\color{red}{pct\_unemploy}} & 0.39 & \textbf{Factor 3:} &  & {\color{brightgreen}{pct\_hous\_owner\_occup}} & 0.20 \\ 
  {\color{red}{pct\_employ\_mfg}} & 0.11 & {\color{patriarch}{pop}} & 0.03 & \textbf{Factor 3:} &  \\ 
  {\color{brightgreen}{pct\_hous\_owner\_occup}} & 0.11 & {\color{red}{pct\_wfarm}} & 0.37 & {\color{red}{pct\_wfarm}} & 0.21 \\ 
  {\color{brightgreen}{pct\_vacant\_boarded}} & 0.86 & {\color{red}{pct\_employ\_prof\_serv}} & 0.63 & {\color{red}{pct\_employ\_prof\_serv}} & 0.42 \\ 
  {\color{brightgreen}{pct\_vacant6up}} & 0.46 & {\color{brightgreen}{own\_hous\_qrange}} & 1.06 & {\color{brightgreen}{own\_hous\_qrange}} & 0.70 \\ 
  {\color{brightgreen}{med\_own\_cost\_pct\_wo}} & 0.36 & {\color{brightgreen}{rent\_qrange}} & 0.72 & {\color{brightgreen}{rent\_qrange}} & 0.35 \\ 
		\hline
\end{tabular}%
}
\caption{Weights $\mathbf{W}_g\mathbf{V}_g$ in $g=1,\dots,6$ for the explanatory variables. The three types of variables:  \textcolor[rgb]{0.5, 0.0, 0.5}{General characteristics}, {\color{red}{Income and Employment}}, \textcolor{brightgreen}{Housing}.}
\label{which}
\end{table}

From Table \ref{which}, it is possible to observe which explanatory variables form the factors and their associated weights $\mathbf{W}_g\mathbf{V}_g$ for each  $g=1,\dots,6$. The latter allows us to evaluate the relative importance of each variable within the corresponding factor for each  $g=1,\dots,6$. The three types of socio-economic variables — General Characteristics, Income and Employment, and Housing — are colored differently for clarity. As expected, depending on the community cluster, each factor is formed by different socio-economic variables which similarly predict the responses. 
In other words, each factor is not equivalent across community clusters. 
Thus, the factor has a different meaning for each community cluster and potentially can act differently on the responses. For example, Factor 1 in Cluster 1 differs from Factor 1 in other clusters, while Factor 1 in Cluster 3 has a similar meaning to Factor 1 in Cluster 6. Generally, the tendency is that the majority of socio-economic variables related to \textcolor{brightgreen}{Housing} cluster together in one or more clusters, such the percentages of vacant houses and vacant houses broadened over 6 months. The same is for socio-economic variables related to {\color{red}{Income}}, such as the percentage of people with social security income and retired workers. Depending on the community cluster, the socio-economic variables related to \textcolor{patriarch}{General characteristics} cluster either with a subset of \textcolor{brightgreen}{Housing} related socio-economic variables or a subset of  \textcolor{red}{Employment} related socio-economic variables.
\begin{table}[ht]
    \centering\footnotesize
   % \arrayrulecolor{black} % Set the color of the table lines to black
    \begin{tabular}{c|c|r|r|r|r|r}
        \hline
        %\rowcolor{gray!30}
   & $b_{1jm}w_{jj}v_{jq}$ & \textbf{Factor 1} & \textbf{Factor 2} & \textbf{Factor 3} \\ 
   \hline
\multirow{3}{*}{Cluster 1} & \textbf{Robberies} & -0.24 & -0.1 & \textcolor{brightgreen}{\textbf{0.4}} \\ 
   & \textbf{Burglaries} & -0.15 & -0.11 & 0.29 \\ 
   & \textbf{Larcenies} & -0.18 & \textcolor{red}{\textbf{-0.26}} & 0.04 \\ 
   \hline
\multirow{3}{*}{Cluster 2} & \textbf{Robberies} & \textcolor{brightgreen}{\textbf{0.23}} & \textcolor{red}{\textbf{-0.34}} & -0.13 \\ 
   & \textbf{Burglaries} & 0.03 & -0.12 & -0.04 \\ 
   & \textbf{Larcenies} & 0.11 & -0.21 & -0.06 \\ 
   \hline
\multirow{3}{*}{Cluster 3} & \textbf{Robberies} & \textcolor{brightgreen}{\textbf{0.31}} & \textcolor{red}{\textbf{-0.37}} & 0.04 \\ 
   & \textbf{Burglaries} & 0.22 & -0.13 & 0.02 \\ 
   & \textbf{Larcenies} & 0.04 & -0.14 & 0.01 \\ 
   \hline
\multirow{3}{*}{Cluster 4} & \textbf{Robberies} & \textcolor{red}{\textbf{-0.31}} & 0.17 & \textcolor{brightgreen}{\textbf{0.62}} \\ 
   & \textbf{Burglaries} & -0.14 & 0.05 & 0.29 \\ 
   & \textbf{Larcenies} & -0.17 & 0.03 & 0.11 \\ 
   \hline
\multirow{3}{*}{Cluster 5} & \textbf{Robberies} & 0.11 & \textcolor{red}{\textbf{-0.19}} & 0 \\ 
   & \textbf{Burglaries} & \textcolor{brightgreen}{\textbf{0.12}} & -0.18 & 0.02 \\ 
   & \textbf{Larcenies} & 0.03 & -0.11 & -0.04 \\ 
   \hline
\multirow{3}{*}{Cluster 6} & \textbf{Robberies} & \textcolor{brightgreen}{\textbf{0.53}} & -0.07 & 0.01 \\ 
   & \textbf{Burglaries} & 0.07 & -0.02 & -0.08 \\ 
   & \textbf{Larcenies} & \textcolor{red}{\textbf{-0.15}} & 0.04 & -0.04 \\ 
   \hline
   \hline
    \end{tabular}
   \tiny{\caption{Regression coefficients $\mathbf{B}_{1g}'\mathbf{W}_g\mathbf{V}_g$ for $g=1,\dots,6$ }
   \label{regred}}
\end{table}

Finally, Table \ref{regred} displays  the regression coefficients in the reduced space computed as $\mathbf{B}_{1g}'\mathbf{W}_g\mathbf{V}_g$ ($g=1,\dots,6$). For each crime, the highest coefficient (by absolute value) is highlighted: {\color{red}{red}} for negative coefficients \textcolor{brightgreen}{ green} for positive coefficients.
From this table, we can observe, for instance, that within Cluster 3, Factor 2 appears to have the strongest negative impact on all responses, while within Cluster 5, Factor 3 appears to have no significant effects on any of the responses. Instead of working with various explanatory variables, we focus on the reduced space of latent factors. In the original space, we need to consider the effects of 16 variables in 6 different groups, which is a challenge for identifying patterns. Additionally, the correlations among the predictors are not taken into account. The disjoint factor structure provides a partition of variables, where the effects are the comprehensive impact of predictors in the groups. We can easily figure out how a factor affects the crime rates, and further investigate the variables related to the factor.\\
In community Cluster 1, both Factor 1 and Factor 2 decrease all responses, whereas a higher Factor 3 significantly raises all the responses, particularly Robberies and Burglaries. It is a reasonable result considering Factor 3 for Cluster 1 includes percentages of unemployment and vacant houses. 
For Cluster 2, Factor 2 and Factor 3 decrease the crime rates overall. Factor 1, including percentages of people retired, unemployed, or with low education or social security income, has the most positive effects on the responses. 
For Cluster 3, Factor 1 has the highest impact on the responses, while Factor 3 also positively influences them less significantly. Factor 2 decreases all crime rates and contains the percentage of the population with farming or self-employed income. Factor 2 also has the percentage of houses used by owners and the percentage of houses vacant for more than 6 months. 
Cluster 4 has the most non-metro counties and the most counties with persistent poverty and low education. Factor 3 significantly raises all the crime rates, and is related to population, retired, unemployed people, education level, social security income, percentage of vacant houses, and median owner cost of houses without a mortgage. Factor 2 also raises crime rates but is not comparable with Factor 3. Factor 1, related to the percentage of people with farming or self-employment income, the percentage of professional employees and the median rent level, impacts the crime rates to the opposite. 
Crime rates of Cluster 5 are mainly influenced by Factor 1 and Factor 2. Factor 1 is more about the level of retirement, unemployment, social security income and rent. A higher Factor 1 indicates higher crime rates. For Factor 2, on the opposite,  a higher value may imply lower crime rates. Factor 2 is related to the group size of workers in manufacturing, vacant housing and the cost of owning a house.
Cluster 6 is mainly associated with Factor 1, of which the effects differ in direction for Robberies and Larcenies. It looks like a higher level of population, low education, unemployment, rent and cost of housing, included in Factor 1, raises Robberies but lowers Larcenies.
\\

\section{Conclusion and discussions}
\label{sec:8}
The proposed MCWDFA extends CWFA in two ways: the multivariate responses and a disjoint factor analyzer. The multivariate assumption allows us to finish the analysis of multiple responses in one run. The disjoint factor assumption provides a more interpretable structure for latent factors. An AECM algorithm was developed for parameter recovery and clustering. A series of simulations were conducted to present the model's performance under various scenarios. Three sets of simulations had different numbers of components and sizes of predictor groups. 
Our model has the ability to figure out the correct latent structures of datasets most of the time under the assessments of BIC and ICL. ARI also measured the clustering accuracies well. Then the performance under different sample sizes and numbers of covariates was tested. The model still presents the ability to recover the true number of components and classify subjects correctly, though the ability decreases as the data set has a higher dimensionality. Then we considered the model's performance on datasets having a not perfect block diagonal matrix. We can see it correctly reveals the latent factors' disjoint structure. Throughout the process, AIC, AIC3, BIC and ICL were used for model assessment but ICL showed a fairly better performance. In the real data analysis part, we thus adopted ICL as the measure to select the optimal model.

We applied the model to the United States' economic crime data at the community level in the 1990s'. The economic crime rates are strongly associated with the local society structure such as population, economic type, education level, housing market and poverty. We expect to uncover the latent disjoint factors under the observed covariates and also the latent subgroups of communities sharing similar societal structures. The proposed method enables us to achieve the goal. However, the factors discovered by the proposed model need further exploration for a better interpretation, though the factors have been associated with much fewer covariates. The proposed model can also be applied to other data sets concerning clustering and latent interpretable factors. Students' scores in multiple subjects with other information and patients' laboratory assessments after treatment with demographics can be potential real scenarios. \\

The empirical findings confirm substantial heterogeneity among United States communities regarding the impact of socio-economic
information on economic crime rates. Additionally, significant variations were observed among socio-economic features in terms of their predictive abilities for crimes across different community clusters. The interesting aspect of the model is that the groups of covariates are associated with factors that are constructed
from disjoint variables (meaning a variable can only belong to one factor), thus
the factors are easily interpretable, as are the associated regression coefficients.
It is worth noting that the factors may potentially change meaning depending
on the cluster.\\
The proposed CWDFA model:
1) Confirms substantial differences among US communities in the impact of
socio-economic factors on economic crime rates, identifying clusters with
similar structures
2) Helps us to discover groups of socio-economic covariates associated with
disjoint factors easier to interpret, as the associated regression
coefficients.\\
Some future directions include improving the model's performance with high-dimension data sets. A sparse structure might be incorporated in the coefficients matrices. As our model considers only continuous numeric data, the extension to categorical and compositional data is of interest, which are widespread data types in the survey data and demographic data. Some model-based hypothesis testing techniques could also be developed.
%%%%%%%%%%%%%%%%%%%%%%%%%%%%%%%%%%%%%%%%%%%%%%%%%%%%%%%%%%%%%%%
%% REFERENCES
%%%%%%%%%%%%%%%%%%%%%%%%%%%%%%%%%%%%%%%%%%%%%%%%%%%%%%%%%%%%%%%%%%%%%%%%%%%%
% \bibliographystyle{apa}
% \bibliography{Cladag2023bib}
%\subsection*{Acknowledgments}
%{\small This research was supported by the National Sciences and Engineering Research Council of Canada grant (2021-03812 - Subedi), and the Canada Research Chair Program (2020-00303 - Subedi).}

%\noindent \textbf{Data Availability Statement}\\
%The dataset used in this study are publicly available via UCI Machine Learning Repository and the Economic Research Service U.S. Department of Agriculture Economic Research Services (https://www.ers.usda.gov/data-products/county-typology-codes/).

%\bibliographystyle{apalike}
%\bibliography{Cladag2023bib}
%%%%%%%%%%%%%%%%%%%%%%%%%%%%%%%%%%%%%%%%%%%%%%%%%%%%%%%%%%%%%%%%%%%%%%%%%%%

\begin{appendices}
\singlespacing
\section{Parameters used for data generation in the simulation study.}
\label{A}

\begin{longtable}{p{.125\textwidth}|p{.875\textwidth}}
%\scriptsize
%\centering
\caption{Pameters used for data generation in Setting 1}\label{simu1:setting1:parameters}\\[-10pt]
\hline 
Parameter&Values\\
\hline
\\
\endfirsthead

\hline 
Parameter&Values\\
\hline 
\endhead
\\
 &Continued on next page...\\ \hline
\endfoot
\hline
\endlastfoot
$\widetilde{\Bv}_1$& $\begin{psmallmatrix}
  -0.16 & -0.75 & 0.75 & -0.54 & -0.58 \\ 
  0.23 & -0.46 & -1.11 & -0.02 & -0.65 \\ 
  -0.21 & -0.70 & -0.67 & -0.86 & -0.07 \\ 
  -0.90 & 0.14 & -1.23 & -0.06 & 0.58 \\ 
  -0.04 & -0.14 & -0.03 & -1.07 & -1.28 \\ 
  -0.08 & -0.56 & -0.40 & -0.91 & 0.03 \\ 
  -0.57 & -1.17 & -0.17 & -1.07 & -0.12 \\ 
  -0.02 & -1.17 & -0.94 & -0.29 & -0.58 \\ 
  -0.62 & 0.06 & -0.30 & -0.87 & -0.40 \\ 
  -0.90 & -0.76 & -1.39 & -0.21 & -0.55 \\ 
  -0.85 & -0.54 & -1.09 & 0.29 & -0.01 \\ 
  -0.00 & -0.69 & -0.40 & -0.58 & -0.25 \\ 
  -0.15 & 0.03 & -0.82 & -0.83 & -0.27 \\ 
  -0.53 & -0.49 & -0.38 & -0.90 & -0.31 \\ 
  -0.81 & -0.81 & 0.08 & -0.86 & -1.17 \\ 
  -0.80 & -0.78 & -0.53 & -1.29 & 0.24 \\ 
  \end{psmallmatrix}$ \\
       &\\ \hspace{2em}
$\widetilde{\Bv}_2$ & $\begin{psmallmatrix}
  0.35 & 0.30 & 1.03 & -1.56 & -0.06 \\ 
  1.09 & -1.26 & -0.80 & -0.61 & 1.34 \\ 
  0.30 & -0.52 & 0.27 & 0.44 & -0.54 \\ 
  0.70 & 1.10 & 0.36 & -1.40 & 1.07 \\ 
  -1.01 & 0.79 & -0.35 & 0.83 & 0.71 \\ 
  -0.37 & -0.52 & -1.11 & -0.79 & 0.08 \\ 
  0.68 & 0.69 & -0.08 & 0.79 & 0.02 \\ 
  -1.69 & -0.16 & 0.24 & 0.72 & 2.02 \\ 
  -0.42 & -0.00 & -0.55 & -0.79 & -0.74 \\ 
  -0.11 & 0.92 & 0.34 & -0.56 & -0.81 \\ 
  -0.82 & 1.09 & 0.20 & 1.11 & 1.10 \\ 
  0.14 & -0.85 & -0.40 & 0.72 & -0.00 \\ 
  -0.78 & 1.48 & -1.02 & 0.14 & 0.14 \\ 
  -0.31 & -0.15 & 0.75 & -0.19 & -0.17 \\ 
  -1.06 & 0.84 & 0.86 & -0.66 & 0.22 \\ 
  -0.13 & -0.06 & 0.41 & -0.51 & -0.27 \\ 
  \end{psmallmatrix}$ \\
     &\\ \hspace{2em}
$\mbox{\boldmath$\Sigma$}_{\ev_1}$ &$ \begin{psmallmatrix}
  2.01 & 1.13 & 0.63 & 0.83 & 0.79 \\ 
  1.13 & 1.83 & 0.97 & 0.81 & 0.50 \\ 
  0.63 & 0.97 & 1.42 & 0.52 & 0.65 \\ 
  0.83 & 0.81 & 0.52 & 1.55 & 0.74 \\ 
  0.79 & 0.50 & 0.65 & 0.74 & 1.00 \\ 
  \end{psmallmatrix}$\\
     &\\ \hspace{2em}
$\mbox{\boldmath$\Sigma$}_{\ev_2}$ & $\begin{psmallmatrix}
  1.79 & 1.46 & 1.18 & 0.30 & 0.61 \\ 
  1.46 & 3.64 & 2.12 & 0.88 & 0.99 \\ 
  1.18 & 2.12 & 1.80 & 0.49 & 0.77 \\ 
  0.30 & 0.88 & 0.49 & 1.00 & 0.04 \\ 
  0.61 & 0.99 & 0.77 & 0.04 & 1.00 \\ 
  \end{psmallmatrix}$\\
     &\\ \hspace{2em}
$\mbox{\boldmath$\mu$}_1$  & $\begin{psmallmatrix}
  -2.57 & -2.95 & -2.54 & -2.85 & -2.40 & -2.60 & -2.69 & -2.01 & -2.86 & -2.57 & -2.09 & -2.40 & -2.52 & -2.59 & -2.60 \\ 
  \end{psmallmatrix}$\\
   &\\ \hspace{2em}
$\mbox{\boldmath$\mu$}_2$  & $\begin{psmallmatrix}
  3.19 & 3.74 & 3.19 & 3.61 & 3.62 & 3.77 & 3.16 & 3.75 & 3.28 & 3.98 & 3.24 & 3.57 & 3.37 & 3.34 & 3.53 \\ 
  \end{psmallmatrix}$\\
   &\\ \hspace{2em}
$\text{diag}(\Wv_{1})$ & $\begin{psmallmatrix}
  1.28 & 0.79 & 1.22 & 0.75 & 1.08 & 0.97 & 1.30 & 1.26 & 1.12 & 1.18 & 1.03 & 0.88 & 0.73 & 0.70 & 1.27 \\ 
  \end{psmallmatrix}$\\
   &\\ \hspace{2em}
$\text{diag}(\Wv_{2})$ & $\begin{psmallmatrix}
  1.64 & 0.88 & 1.62 & 0.80 & 1.18 & 0.93 & 1.44 & 0.72 & 0.83 & 1.47 & 0.94 & 1.83 & 0.95 & 1.90 & 1.70 \\ 
  \end{psmallmatrix}$\\
   &\\ \hspace{2em}
$\text{diag}(\mbox{\boldmath$\Psi$}_1)$ & $\begin{psmallmatrix}
  0.96 & 0.48 & 0.72 & 0.85 & 0.29 & 0.02 & 0.28 & 0.29 & 0.24 & 0.69 & 0.87 & 0.57 & 0.29 & 0.69 & 0.38 \\ 
  \end{psmallmatrix}$\\
   &\\ \hspace{2em}
$\text{diag}(\mbox{\boldmath$\Psi$}_2)$ & $\begin{psmallmatrix}
  0.50 & 0.67 & 0.32 & 0.14 & 0.77 & 0.19 & 0.18 & 0.09 & 0.67 & 0.99 & 0.09 & 0.28 & 0.69 & 0.34 & 0.51 \\ 
  \end{psmallmatrix}$\\
\end{longtable}

\begin{longtable}{p{.125\textwidth}|p{.875\textwidth}}
%\scriptsize
%\centering
\caption{Pameters used for data generation in Setting 2.}\label{simu1:setting2:parameters}\\[-10pt]
\hline 
Parameter&Values\\
\hline
\\
\endfirsthead

\hline 
Parameter&Values\\
\hline 
\endhead
 &Continued on next page...\\ \hline
\endfoot
\hline
\endlastfoot
$\widetilde{\Bv}_1$& $\begin{psmallmatrix}
  -0.16 & -0.75 & 0.75 & -0.54 & -0.58 \\ 
  0.23 & -0.46 & -1.11 & -0.02 & -0.65 \\ 
  -0.21 & -0.70 & -0.67 & -0.86 & -0.07 \\ 
  -0.90 & 0.14 & -1.23 & -0.06 & 0.58 \\ 
  -0.04 & -0.14 & -0.03 & -1.07 & -1.28 \\ 
  -0.08 & -0.56 & -0.40 & -0.91 & 0.03 \\ 
  -0.57 & -1.17 & -0.17 & -1.07 & -0.12 \\ 
  -0.02 & -1.17 & -0.94 & -0.29 & -0.58 \\ 
  -0.62 & 0.06 & -0.30 & -0.87 & -0.40 \\ 
  -0.90 & -0.76 & -1.39 & -0.21 & -0.55 \\ 
  -0.85 & -0.54 & -1.09 & 0.29 & -0.01 \\ 
  -0.00 & -0.69 & -0.40 & -0.58 & -0.25 \\ 
  -0.15 & 0.03 & -0.82 & -0.83 & -0.27 \\ 
  -0.53 & -0.49 & -0.38 & -0.90 & -0.31 \\ 
  -0.81 & -0.81 & 0.08 & -0.86 & -1.17 \\ 
  -0.80 & -0.78 & -0.53 & -1.29 & 0.24 \\ 
  \end{psmallmatrix}$ \\
     &\\ \hspace{2em}
$\widetilde{\Bv}_2$ & $\begin{psmallmatrix}
  0.35 & 0.30 & 1.03 & -1.56 & -0.06 \\ 
  1.09 & -1.26 & -0.80 & -0.61 & 1.34 \\ 
  0.30 & -0.52 & 0.27 & 0.44 & -0.54 \\ 
  0.70 & 1.10 & 0.36 & -1.40 & 1.07 \\ 
  -1.01 & 0.79 & -0.35 & 0.83 & 0.71 \\ 
  -0.37 & -0.52 & -1.11 & -0.79 & 0.08 \\ 
  0.68 & 0.69 & -0.08 & 0.79 & 0.02 \\ 
  -1.69 & -0.16 & 0.24 & 0.72 & 2.02 \\ 
  -0.42 & -0.00 & -0.55 & -0.79 & -0.74 \\ 
  -0.11 & 0.92 & 0.34 & -0.56 & -0.81 \\ 
  -0.82 & 1.09 & 0.20 & 1.11 & 1.10 \\ 
  0.14 & -0.85 & -0.40 & 0.72 & -0.00 \\ 
  -0.78 & 1.48 & -1.02 & 0.14 & 0.14 \\ 
  -0.31 & -0.15 & 0.75 & -0.19 & -0.17 \\ 
  -1.06 & 0.84 & 0.86 & -0.66 & 0.22 \\ 
  -0.13 & -0.06 & 0.41 & -0.51 & -0.27 \\ 
  \end{psmallmatrix}$ \\
     &\\ \hspace{2em}
$\mbox{\boldmath$\Sigma$}_{\ev_1}$ &$ \begin{psmallmatrix}
  2.01 & 1.13 & 0.63 & 0.83 & 0.79 \\ 
  1.13 & 1.83 & 0.97 & 0.81 & 0.50 \\ 
  0.63 & 0.97 & 1.42 & 0.52 & 0.65 \\ 
  0.83 & 0.81 & 0.52 & 1.55 & 0.74 \\ 
  0.79 & 0.50 & 0.65 & 0.74 & 1.00 \\ 
  \end{psmallmatrix}$\\
     &\\ \hspace{2em}
$\mbox{\boldmath$\Sigma$}_{\ev_2}$ & $\begin{psmallmatrix}
  1.79 & 1.46 & 1.18 & 0.30 & 0.61 \\ 
  1.46 & 3.64 & 2.12 & 0.88 & 0.99 \\ 
  1.18 & 2.12 & 1.80 & 0.49 & 0.77 \\ 
  0.30 & 0.88 & 0.49 & 1.00 & 0.04 \\ 
  0.61 & 0.99 & 0.77 & 0.04 & 1.00 \\ 
  \end{psmallmatrix}$\\
     &\\ \hspace{2em}
$\mbox{\boldmath$\mu$}_1$  & $\begin{psmallmatrix}
  -2.57 & -2.95 & -2.54 & -2.85 & -2.40 & -2.60 & -2.69 & -2.01 & -2.86 & -2.57 & -2.09 & -2.40 & -2.52 & -2.59 & -2.60 \\ 
  \end{psmallmatrix}$\\
   &\\ \hspace{2em}
$\mbox{\boldmath$\mu$}_2$  & $\begin{psmallmatrix}
  3.19 & 3.74 & 3.19 & 3.61 & 3.62 & 3.77 & 3.16 & 3.75 & 3.28 & 3.98 & 3.24 & 3.57 & 3.37 & 3.34 & 3.53 \\ 
  \end{psmallmatrix}$\\
   &\\ \hspace{2em}
$\text{diag}(\Wv_{1})$ & $\begin{psmallmatrix}
  1.28 & 0.79 & 1.22 & 0.75 & 1.08 & 0.97 & 1.30 & 1.26 & 1.12 & 1.18 & 1.03 & 0.88 & 0.73 & 0.70 & 1.27 \\ 
  \end{psmallmatrix}$\\
   &\\ \hspace{2em}
$\text{diag}(\Wv_{2})$ & $\begin{psmallmatrix}
  1.64 & 0.88 & 1.62 & 0.80 & 1.18 & 0.93 & 1.44 & 0.72 & 0.83 & 1.47 & 0.94 & 1.83 & 0.95 & 1.90 & 1.70 \\ 
  \end{psmallmatrix}$\\
   &\\ \hspace{2em}
$\text{diag}(\mbox{\boldmath$\Psi$}_1)$ & $\begin{psmallmatrix}
  0.96 & 0.48 & 0.72 & 0.85 & 0.29 & 0.02 & 0.28 & 0.29 & 0.24 & 0.69 & 0.87 & 0.57 & 0.29 & 0.69 & 0.38 \\ 
  \end{psmallmatrix}$\\
   &\\ \hspace{2em}
$\text{diag}(\mbox{\boldmath$\Psi$}_2)$ & $\begin{psmallmatrix}
  0.50 & 0.67 & 0.32 & 0.14 & 0.77 & 0.19 & 0.18 & 0.09 & 0.67 & 0.99 & 0.09 & 0.28 & 0.69 & 0.34 & 0.51 \\ 
  \end{psmallmatrix}$\\
\end{longtable}

\begin{longtable}{p{.125\textwidth}|p{.875\textwidth}}
%\begin{longtable}{c{.1\textwidth}c{.9\textwidth}}
%\scriptsize
%\centering
\caption{Pameters used for data generation in Setting 3.}\label{simu1:setting3:parameters}\\[-10pt]
\hline 
Parameter&Values\\
\hline
\\
\endfirsthead

\hline 
Parameter&Values\\
\hline 
\\
\endhead
 &Continued on next page...\\ \hline
\endfoot
\hline
\endlastfoot
$\widetilde{\Bv}_1$& $\begin{psmallmatrix}
  -0.55 & -0.71 & -0.31 & -0.15 & -1.11 & 0.44 & -0.15 & -1.10 & -0.51 & -0.12 \\ 
  0.09 & -0.55 & 1.14 & -0.11 & -0.50 & -0.82 & -0.09 & -0.24 & -0.87 & 0.27 \\ 
  0.68 & -0.18 & -0.63 & 0.05 & -0.57 & 0.33 & -1.53 & -0.55 & -1.22 & 0.43 \\ 
  0.17 & -0.47 & -0.57 & -0.49 & 0.35 & -0.07 & -0.20 & -0.68 & 0.44 & -1.60 \\ 
  -0.67 & 0.31 & -0.83 & 0.06 & 1.03 & -0.15 & -0.02 & -0.11 & -0.93 & 0.50 \\ 
  0.38 & 0.28 & 0.29 & -1.04 & -0.93 & -0.66 & -0.35 & 0.42 & 0.30 & -0.12 \\ 
  -0.01 & -0.09 & -0.36 & -0.56 & 0.40 & -0.99 & -0.01 & -0.91 & -0.33 & -1.08 \\ 
  -0.24 & -1.06 & -0.02 & -1.04 & 0.31 & -0.30 & -0.15 & 0.17 & 0.69 & 0.31 \\ 
  0.35 & -0.76 & -0.84 & -0.07 & -0.43 & -1.27 & 0.41 & -0.37 & -0.36 & 0.66 \\ 
  -0.51 & 0.10 & 0.06 & -0.40 & 0.09 & -0.44 & 0.04 & -0.66 & -0.42 & -1.15 \\ 
  -0.70 & -0.60 & -1.19 & -0.09 & -0.31 & 0.64 & -0.27 & -0.87 & -0.87 & -1.46 \\ 
  -0.83 & -0.07 & -0.66 & 0.51 & 0.29 & -0.64 & 0.49 & 0.23 & 0.30 & 0.06 \\ 
  0.17 & -0.29 & -0.21 & -0.16 & 0.21 & 0.26 & -0.81 & 0.27 & -0.09 & -0.55 \\ 
  -0.01 & 0.20 & -0.74 & -0.47 & 0.11 & -0.28 & 0.38 & -0.85 & -0.24 & -0.83 \\ 
  -0.52 & 0.00 & -0.26 & -0.55 & 0.07 & 0.27 & -0.01 & -0.28 & 0.11 & 0.38 \\ 
  -0.77 & -0.61 & 0.56 & -0.78 & -1.14 & -0.82 & -0.40 & 0.08 & -0.58 & 0.26 \\ 
  \end{psmallmatrix}$ \\
     &\\ \hspace{2em}
$\widetilde{\Bv}_2$ & $\begin{psmallmatrix}
  0.13 & 1.01 & 0.76 & -0.37 & 0.69 & -1.40 & -0.22 & 0.32 & 1.15 & -1.15 \\ 
  0.47 & -0.41 & 1.31 & -0.96 & 0.66 & -0.85 & -0.49 & 0.52 & 0.54 & 1.21 \\ 
  0.37 & 0.10 & 0.08 & 0.07 & -0.77 & -0.71 & -0.12 & 0.93 & 0.71 & 0.04 \\ 
  0.44 & -0.14 & 0.65 & -0.09 & -1.10 & -0.78 & 0.74 & -0.69 & 1.13 & -0.01 \\ 
  -0.31 & -0.86 & -0.42 & -1.09 & 0.29 & 1.15 & 0.14 & -1.03 & -0.63 & -0.77 \\ 
  1.08 & 1.12 & 1.20 & -0.57 & -0.36 & -0.01 & -0.26 & 1.09 & -0.72 & -1.11 \\ 
  -0.12 & 0.35 & -0.92 & 0.22 & -0.26 & 1.24 & -0.35 & 0.50 & -0.73 & -0.90 \\ 
  -0.85 & -0.23 & -0.44 & 0.18 & -0.15 & -0.18 & 0.34 & -0.54 & 0.76 & -0.78 \\ 
  -0.69 & 1.51 & -1.23 & 0.52 & -0.03 & 0.34 & 0.48 & -0.83 & 1.09 & -1.67 \\ 
  0.72 & -0.54 & 0.47 & -0.54 & -0.70 & -0.89 & -0.70 & -2.19 & -0.38 & -0.05 \\ 
  -0.13 & 0.78 & 0.80 & 0.01 & -1.54 & 0.81 & -0.10 & -0.58 & 0.28 & 0.37 \\ 
  -0.43 & 0.66 & 0.21 & -1.22 & 1.83 & 0.11 & 0.25 & 0.43 & 0.02 & 0.91 \\ 
  -1.04 & 0.23 & 0.46 & 0.78 & -0.03 & -0.69 & 1.49 & -1.21 & 0.16 & 0.06 \\ 
  0.65 & -0.48 & -0.44 & -0.76 & -0.27 & 0.95 & -0.86 & 0.48 & 0.30 & -0.05 \\ 
  0.06 & -0.71 & -1.86 & 0.27 & -0.11 & -1.36 & -0.11 & 0.02 & -1.23 & -2.24 \\ 
  0.37 & -1.28 & 0.27 & -0.46 & 0.03 & 0.82 & -1.38 & -1.03 & -0.34 & -0.63 \\ 
  \end{psmallmatrix}$ \\
     &\\ \hspace{2em}
$\widetilde{\Bv}_3$ & $\begin{psmallmatrix}
  -1.74 & 1.52 & -1.76 & 0.83 & -0.04 & 0.67 & 0.84 & -0.13 & 0.10 & 1.26 \\ 
  -0.16 & -1.59 & 0.82 & 0.99 & 1.08 & -1.87 & 0.17 & 0.07 & -0.86 & -0.27 \\ 
  -0.42 & 0.06 & 0.44 & 0.02 & 0.41 & 0.44 & 0.14 & -1.40 & -0.42 & -1.62 \\ 
  -0.11 & 0.30 & -1.21 & 0.02 & 0.29 & 1.17 & -1.36 & -0.73 & 0.50 & -1.33 \\ 
  -0.44 & 0.84 & 0.13 & -0.36 & 1.62 & 0.13 & -1.53 & 1.69 & 0.35 & 1.08 \\ 
  -0.56 & -0.06 & 1.10 & -0.33 & -0.98 & 0.29 & 0.12 & 0.49 & -1.15 & -0.47 \\ 
  -0.41 & -1.60 & -0.68 & -0.73 & -0.49 & -1.28 & -0.05 & 1.77 & -0.97 & 0.64 \\ 
  0.20 & 1.08 & -0.29 & 0.71 & -0.44 & -1.28 & 0.42 & 1.31 & -0.25 & 0.04 \\ 
  0.57 & 0.43 & 0.66 & -1.42 & -0.15 & -1.82 & 0.76 & -1.13 & 0.38 & -0.22 \\ 
  -0.69 & -0.00 & -1.14 & 0.21 & 0.04 & -0.02 & -0.47 & -0.26 & -1.74 & 2.28 \\ 
  -1.40 & 0.64 & -0.37 & -1.60 & -2.22 & -1.40 & 0.36 & 0.35 & -0.38 & -0.84 \\ 
  -1.33 & -0.90 & -1.25 & -1.77 & 0.65 & -0.77 & 0.49 & -0.72 & -1.39 & -0.61 \\ 
  -0.43 & 0.46 & -0.28 & 0.16 & -2.36 & 1.39 & -2.95 & -0.49 & 0.58 & -1.38 \\ 
  0.69 & 0.76 & 0.25 & -1.51 & -0.91 & 1.47 & -0.50 & 1.01 & -0.65 & 0.54 \\ 
  0.19 & -0.13 & 1.01 & 0.39 & -2.12 & 0.52 & 0.29 & -0.45 & -1.41 & 1.05 \\ 
  0.03 & 0.71 & -0.41 & 0.22 & -0.35 & -0.19 & 0.55 & -1.72 & -1.46 & 0.13 \\ 
  \end{psmallmatrix}$ \\
     &\\ \hspace{2em}
$\mbox{\boldmath$\Sigma$}_{\ev_1}$ &$ \begin{psmallmatrix}
  5.02 & 3.34 & 3.40 & 2.44 & 1.87 & 2.49 & 1.60 & 1.70 & 1.09 & 0.92 \\ 
  3.34 & 3.49 & 2.53 & 1.59 & 1.87 & 1.91 & 0.99 & 0.76 & 0.84 & 0.64 \\ 
  3.40 & 2.53 & 4.13 & 2.19 & 2.17 & 2.10 & 1.82 & 1.56 & 0.31 & 0.74 \\ 
  2.44 & 1.59 & 2.19 & 2.84 & 0.90 & 1.37 & 0.98 & 1.66 & 0.44 & 0.76 \\ 
  1.87 & 1.87 & 2.17 & 0.90 & 2.77 & 1.81 & 1.06 & 0.23 & 0.19 & 0.02 \\ 
  2.49 & 1.91 & 2.10 & 1.37 & 1.81 & 2.73 & 1.49 & 0.78 & 1.01 & 0.10 \\ 
  1.60 & 0.99 & 1.82 & 0.98 & 1.06 & 1.49 & 1.77 & 0.99 & 0.66 & 0.47 \\ 
  1.70 & 0.76 & 1.56 & 1.66 & 0.23 & 0.78 & 0.99 & 1.77 & 0.36 & 0.88 \\ 
  1.09 & 0.84 & 0.31 & 0.44 & 0.19 & 1.01 & 0.66 & 0.36 & 1.11 & 0.33 \\ 
  0.92 & 0.64 & 0.74 & 0.76 & 0.02 & 0.10 & 0.47 & 0.88 & 0.33 & 1.00 \\ 
  \end{psmallmatrix}$\\
     &\\ \hspace{2em}
$\mbox{\boldmath$\Sigma$}_{\ev_2}$ & $\begin{psmallmatrix}
  4.55 & 3.12 & 2.18 & 2.65 & 2.37 & 2.82 & 1.90 & 2.11 & 1.53 & 0.79 \\ 
  3.12 & 4.06 & 3.03 & 2.32 & 2.40 & 1.92 & 0.91 & 1.18 & 0.94 & 0.72 \\ 
  2.18 & 3.03 & 3.65 & 1.92 & 2.19 & 1.48 & 0.65 & 0.59 & 0.32 & 0.06 \\ 
  2.65 & 2.32 & 1.92 & 3.13 & 1.71 & 2.34 & 1.54 & 2.08 & 1.52 & 0.83 \\ 
  2.37 & 2.40 & 2.19 & 1.71 & 3.20 & 2.48 & 1.63 & 1.25 & 0.62 & 0.69 \\ 
  2.82 & 1.92 & 1.48 & 2.34 & 2.48 & 3.05 & 1.98 & 1.99 & 1.28 & 0.94 \\ 
  1.90 & 0.91 & 0.65 & 1.54 & 1.63 & 1.98 & 2.25 & 1.77 & 1.00 & 0.55 \\ 
  2.11 & 1.18 & 0.59 & 2.08 & 1.25 & 1.99 & 1.77 & 2.58 & 1.66 & 0.83 \\ 
  1.53 & 0.94 & 0.32 & 1.52 & 0.62 & 1.28 & 1.00 & 1.66 & 1.74 & 0.86 \\ 
  0.79 & 0.72 & 0.06 & 0.83 & 0.69 & 0.94 & 0.55 & 0.83 & 0.86 & 1.00 \\ 
  \end{psmallmatrix}$\\
     &\\ \hspace{2em}
$\mbox{\boldmath$\Sigma$}_{\ev_3}$ & $\begin{psmallmatrix}
  2.92 & 1.73 & 1.24 & 1.06 & 0.68 & 1.10 & 0.91 & 1.21 & 0.87 & 0.63 \\ 
  1.73 & 3.65 & 1.65 & 2.20 & 0.91 & 1.52 & 1.12 & 0.27 & 0.16 & 0.14 \\ 
  1.24 & 1.65 & 2.12 & 1.21 & 0.97 & 1.48 & 0.31 & 0.54 & 0.32 & 0.01 \\ 
  1.06 & 2.20 & 1.21 & 3.03 & 1.09 & 2.21 & 1.38 & 1.22 & 0.31 & 0.23 \\ 
  0.68 & 0.91 & 0.97 & 1.09 & 1.46 & 1.13 & 0.58 & 0.48 & 0.26 & 0.08 \\ 
  1.10 & 1.52 & 1.48 & 2.21 & 1.13 & 2.89 & 1.27 & 1.39 & 0.81 & 0.15 \\ 
  0.91 & 1.12 & 0.31 & 1.38 & 0.58 & 1.27 & 1.91 & 1.07 & 0.16 & 0.91 \\ 
  1.21 & 0.27 & 0.54 & 1.22 & 0.48 & 1.39 & 1.07 & 2.18 & 0.81 & 0.81 \\ 
  0.87 & 0.16 & 0.32 & 0.31 & 0.26 & 0.81 & 0.16 & 0.81 & 1.01 & 0.11 \\ 
  0.63 & 0.14 & 0.01 & 0.23 & 0.08 & 0.15 & 0.91 & 0.81 & 0.11 & 1.00 \\ 
  \end{psmallmatrix}$\\
     &\\ \hspace{2em}
$\mbox{\boldmath$\mu$}_1$  & $\begin{psmallmatrix}
  -0.61 & -0.07 & 0.08 & 0.08 & -0.68 & -0.69 & 0.06 & -0.67 & 0.23 & -0.67 & 0.23 & -0.20 & -0.57 & 0.08 & -0.02 \\ 
  \end{psmallmatrix}$\\
   &\\ \hspace{2em}
$\mbox{\boldmath$\mu$}_2$  & $\begin{psmallmatrix}
  1.71 & 0.95 & 0.80 & 1.46 & 1.41 & 1.20 & 0.95 & 0.96 & 1.28 & 1.61 & 1.17 & 0.77 & 1.50 & 0.81 & 1.17 \\ 
  \end{psmallmatrix}$\\
   &\\ \hspace{2em}
$\mbox{\boldmath$\mu$}_3$  & $\begin{psmallmatrix}
  2.82 & 2.55 & 2.87 & 2.41 & 2.39 & 3.14 & 3.13 & 2.67 & 3.07 & 2.55 & 2.50 & 2.91 & 2.28 & 2.83 & 2.90 \\ 
  \end{psmallmatrix}$\\
   &\\ \hspace{2em}
$\text{diag}(\Wv_{1})$ & $\begin{psmallmatrix}
  1.28 & 0.79 & 1.22 & 0.75 & 1.08 & 0.97 & 1.30 & 1.26 & 1.12 & 1.18 & 1.03 & 0.88 & 0.73 & 0.70 & 1.27 \\ 
  \end{psmallmatrix}$\\
   &\\ \hspace{2em}
$\text{diag}(\Wv_{2})$ & $\begin{psmallmatrix}
  1.45 & 0.28 & 1.41 & 0.16 & 0.74 & 0.35 & 1.14 & 0.02 & 0.20 & 1.19 & 0.37 & 1.74 & 0.39 & 1.85 & 1.54 \\ 
  \end{psmallmatrix}$\\
   &\\ \hspace{2em}
$\text{diag}(\Wv_{3})$ & $\begin{psmallmatrix}
  1.00 & 1.00 & 1.00 & 1.00 & 1.00 & 1.00 & 1.00 & 1.00 & 1.00 & 1.00 & 1.00 & 1.00 & 1.00 & 1.00 & 1.00 \\ 
  \end{psmallmatrix}$\\
   &\\ \hspace{2em}
$\text{diag}(\mbox{\boldmath$\Psi$}_1)$ & $\begin{psmallmatrix}
  0.61 & 0.36 & 0.21 & 0.69 & 0.07 & 0.52 & 0.19 & 0.78 & 0.56 & 0.16 & 0.20 & 0.67 & 0.65 & 0.18 & 0.86 \\ 
  \end{psmallmatrix}$\\
   &\\ \hspace{2em}
$\text{diag}(\mbox{\boldmath$\Psi$}_2)$ & $\begin{psmallmatrix}
  0.68 & 0.11 & 0.85 & 0.42 & 0.34 & 0.87 & 0.38 & 0.61 & 0.47 & 0.29 & 0.54 & 0.23 & 0.70 & 0.32 & 0.53 \\ 
  \end{psmallmatrix}$\\
   &\\ \hspace{2em}
$\text{diag}(\mbox{\boldmath$\Psi$}_3)$ & $\begin{psmallmatrix}
  0.37 & 0.88 & 0.21 & 0.81 & 0.75 & 0.05 & 0.23 & 0.96 & 0.97 & 0.24 & 0.16 & 0.77 & 0.80 & 0.56 & 0.90 \\ 
  \end{psmallmatrix}$
\end{longtable}

\section{Component-specific parameters recovery}
\label{B}

\begin{longtable}{p{.125\textwidth}|p{.875\textwidth}}
%\scriptsize
%\centering
\caption{The average MSE of parameters when the correct model was selected (ICL) in Setting 1.}\label{simu1:setting1:MSE}\\
\hline 
Parameter&Mean squared Error\\
\hline
\\
\endfirsthead

\hline 
Parameter&Mean squared Error\\
\hline 
\\
\endhead
 &Continued on next page...\\ \hline
\endfoot
\hline
\endlastfoot
$\widetilde{\Bv}_1$& $\begin{psmallmatrix}
  0.23 & 0.18 & 0.11 & 0.19 & 0.14 \\ 
  0.01 & 0.01 & 0.00 & 0.01 & 0.00 \\ 
  0.01 & 0.01 & 0.01 & 0.01 & 0.01 \\ 
  0.01 & 0.01 & 0.01 & 0.01 & 0.00 \\ 
  0.01 & 0.01 & 0.01 & 0.01 & 0.00 \\ 
  0.02 & 0.02 & 0.01 & 0.01 & 0.01 \\ 
  0.06 & 0.05 & 0.05 & 0.06 & 0.03 \\ 
  0.02 & 0.02 & 0.01 & 0.02 & 0.01 \\ 
  0.02 & 0.01 & 0.01 & 0.01 & 0.01 \\ 
  0.02 & 0.02 & 0.01 & 0.01 & 0.01 \\ 
  0.01 & 0.01 & 0.01 & 0.01 & 0.00 \\ 
  0.01 & 0.01 & 0.01 & 0.01 & 0.01 \\ 
  0.01 & 0.01 & 0.01 & 0.01 & 0.01 \\ 
  0.02 & 0.02 & 0.01 & 0.01 & 0.01 \\ 
  0.01 & 0.01 & 0.01 & 0.01 & 0.01 \\ 
  0.01 & 0.01 & 0.01 & 0.01 & 0.01 \\ 
  \end{psmallmatrix}$ \\
   &\\ \hspace{2em}
$\widetilde{\Bv}_2$ & $\begin{psmallmatrix}
  1.41 & 2.87 & 1.28 & 0.82 & 0.71 \\ 
  0.01 & 0.03 & 0.01 & 0.01 & 0.01 \\ 
  0.01 & 0.03 & 0.01 & 0.01 & 0.01 \\ 
  0.03 & 0.05 & 0.02 & 0.01 & 0.01 \\ 
  0.05 & 0.11 & 0.06 & 0.04 & 0.03 \\ 
  0.01 & 0.03 & 0.01 & 0.01 & 0.01 \\ 
  0.04 & 0.09 & 0.04 & 0.02 & 0.02 \\ 
  0.03 & 0.08 & 0.04 & 0.02 & 0.02 \\ 
  0.09 & 0.18 & 0.09 & 0.05 & 0.05 \\ 
  0.01 & 0.03 & 0.02 & 0.01 & 0.01 \\ 
  0.01 & 0.02 & 0.01 & 0.01 & 0.01 \\ 
  0.11 & 0.22 & 0.11 & 0.04 & 0.04 \\ 
  0.02 & 0.05 & 0.03 & 0.01 & 0.02 \\ 
  0.01 & 0.03 & 0.01 & 0.01 & 0.01 \\ 
  0.02 & 0.04 & 0.02 & 0.01 & 0.01 \\ 
  0.01 & 0.03 & 0.01 & 0.01 & 0.01 \\ 
  \end{psmallmatrix}$ \\
   &\\ \hspace{2em}
$\mbox{\boldmath$\Sigma$}_{\ev_1}$ &$ \begin{psmallmatrix}
  0.03 & 0.02 & 0.01 & 0.01 & 0.01 \\ 
  0.02 & 0.03 & 0.02 & 0.01 & 0.01 \\ 
  0.01 & 0.02 & 0.02 & 0.01 & 0.01 \\ 
  0.01 & 0.01 & 0.01 & 0.02 & 0.01 \\ 
  0.01 & 0.01 & 0.01 & 0.01 & 0.01 \\ 
  \end{psmallmatrix}$\\
   &\\ \hspace{2em}
$\mbox{\boldmath$\Sigma$}_{\ev_2}$ & $\begin{psmallmatrix}
  0.05 & 0.05 & 0.03 & 0.01 & 0.01 \\ 
  0.05 & 0.20 & 0.07 & 0.03 & 0.04 \\ 
  0.03 & 0.07 & 0.04 & 0.01 & 0.02 \\ 
  0.01 & 0.03 & 0.01 & 0.02 & 0.00 \\ 
  0.01 & 0.04 & 0.02 & 0.00 & 0.02 \\ 
  \end{psmallmatrix}$\\
   &\\ \hspace{2em}
$\mbox{\boldmath$\mu$}_1$  & $\begin{psmallmatrix}
  0.01 & 0.00 & 0.01 & 0.01 & 0.01 & 0.00 & 0.01 & 0.01 & 0.01 & 0.01 & 0.01 & 0.00 & 0.00 & 0.00 & 0.01 \\ 
  \end{psmallmatrix}$\\
   &\\ \hspace{2em}
$\mbox{\boldmath$\mu$}_2$  & $\begin{psmallmatrix}
  0.01 & 0.01 & 0.01 & 0.00 & 0.01 & 0.01 & 0.01 & 0.00 & 0.01 & 0.01 & 0.01 & 0.02 & 0.01 & 0.02 & 0.02 \\ 
  \end{psmallmatrix}$\\
 &\\ \hspace{2em}
$\text{diag}(\Wv_{1})$ & $\begin{psmallmatrix}
  0.02 & 0.00 & 0.00 & 0.00 & 0.00 & 0.00 & 0.00 & 0.02 & 0.00 & 0.02 & 0.00 & 0.01 & 0.01 & 0.00 & 0.00 \\ 
  \end{psmallmatrix}$\\
 &\\ \hspace{2em}
$\text{diag}(\Wv_{2})$ & $\begin{psmallmatrix}
  0.03 & 0.01 & 0.06 & 0.00 & 0.01 & 0.01 & 0.01 & 0.01 & 0.00 & 0.03 & 0.02 & 0.01 & 0.03 & 0.09 & 0.03 \\ 
  \end{psmallmatrix}$\\
 &\\ \hspace{2em}
$\text{diag}(\mbox{\boldmath$\Psi$}_1)$ & $\begin{psmallmatrix}
  0.03 & 0.00 & 0.00 & 0.00 & 0.00 & 0.00 & 0.00 & 0.02 & 0.00 & 0.03 & 0.00 & 0.01 & 0.00 & 0.00 & 0.00 \\ 
  \end{psmallmatrix}$\\
 &\\ \hspace{2em}
$\text{diag}(\mbox{\boldmath$\Psi$}_2)$ & $\begin{psmallmatrix}
  0.08 & 0.01 & 0.22 & 0.00 & 0.01 & 0.01 & 0.00 & 0.01 & 0.00 & 0.06 & 0.02 & 0.00 & 0.04 & 0.40 & 0.07 \\ 
  \end{psmallmatrix}
$\\

\end{longtable}

\begin{longtable}{p{.125\textwidth}|p{.875\textwidth}}
%\scriptsize
%\centering
\caption{The average MSE of parameters when the correct model was selected (ICL) in Setting 2.}\label{simu1:setting2:MSE}\\
\hline 
Parameter&Mean squared Error\\
\hline
\\
\endfirsthead

\hline 
Parameter&Mean squared Error\\
\hline 
\\
\endhead
 &Continued on next page...\\ \hline
\endfoot
\hline
\endlastfoot
$\widetilde{\Bv}_1$& $\begin{psmallmatrix}
  0.21 & 0.16 & 0.12 & 0.16 & 0.13 \\ 
  0.01 & 0.01 & 0.00 & 0.01 & 0.00 \\ 
  0.01 & 0.01 & 0.01 & 0.01 & 0.01 \\ 
  0.01 & 0.01 & 0.01 & 0.01 & 0.00 \\ 
  0.01 & 0.01 & 0.01 & 0.01 & 0.00 \\ 
  0.02 & 0.02 & 0.01 & 0.02 & 0.01 \\ 
  0.05 & 0.04 & 0.03 & 0.05 & 0.02 \\ 
  0.02 & 0.02 & 0.01 & 0.02 & 0.01 \\ 
  0.01 & 0.01 & 0.01 & 0.01 & 0.01 \\ 
  0.02 & 0.02 & 0.02 & 0.02 & 0.01 \\ 
  0.01 & 0.01 & 0.01 & 0.01 & 0.00 \\ 
  0.01 & 0.01 & 0.01 & 0.01 & 0.00 \\ 
  0.01 & 0.01 & 0.01 & 0.01 & 0.00 \\ 
  0.02 & 0.02 & 0.01 & 0.01 & 0.01 \\ 
  0.01 & 0.01 & 0.01 & 0.01 & 0.01 \\ 
  0.01 & 0.01 & 0.01 & 0.01 & 0.01 \\ 
  \end{psmallmatrix}$ \\
   &\\ \hspace{2em}
$\widetilde{\Bv}_2$ & $\begin{psmallmatrix}
  1.41 & 2.68 & 1.18 & 0.80 & 0.69 \\ 
  0.01 & 0.03 & 0.01 & 0.01 & 0.01 \\ 
  0.01 & 0.02 & 0.01 & 0.01 & 0.01 \\ 
  0.03 & 0.06 & 0.03 & 0.01 & 0.01 \\ 
  0.05 & 0.11 & 0.06 & 0.04 & 0.03 \\ 
  0.01 & 0.03 & 0.01 & 0.01 & 0.01 \\ 
  0.04 & 0.08 & 0.04 & 0.02 & 0.02 \\ 
  0.03 & 0.08 & 0.04 & 0.02 & 0.02 \\ 
  0.10 & 0.18 & 0.09 & 0.05 & 0.05 \\ 
  0.01 & 0.03 & 0.02 & 0.01 & 0.01 \\ 
  0.01 & 0.02 & 0.01 & 0.01 & 0.00 \\ 
  0.11 & 0.21 & 0.11 & 0.04 & 0.04 \\ 
  0.02 & 0.05 & 0.03 & 0.01 & 0.01 \\ 
  0.01 & 0.03 & 0.01 & 0.01 & 0.01 \\ 
  0.02 & 0.04 & 0.02 & 0.01 & 0.01 \\ 
  0.01 & 0.03 & 0.01 & 0.01 & 0.01 \\ 
  \end{psmallmatrix}$ \\
   &\\ \hspace{2em}
$\mbox{\boldmath$\Sigma$}_{\ev_1}$ &$ \begin{psmallmatrix}
  0.03 & 0.02 & 0.01 & 0.01 & 0.01 \\ 
  0.02 & 0.03 & 0.02 & 0.01 & 0.01 \\ 
  0.01 & 0.02 & 0.02 & 0.01 & 0.01 \\ 
  0.01 & 0.01 & 0.01 & 0.02 & 0.01 \\ 
  0.01 & 0.01 & 0.01 & 0.01 & 0.01 \\ 
  \end{psmallmatrix}$\\
   &\\ \hspace{2em}
$\mbox{\boldmath$\Sigma$}_{\ev_2}$ & $\begin{psmallmatrix}
  0.05 & 0.05 & 0.03 & 0.01 & 0.01 \\ 
  0.05 & 0.20 & 0.07 & 0.03 & 0.04 \\ 
  0.03 & 0.07 & 0.04 & 0.01 & 0.02 \\ 
  0.01 & 0.03 & 0.01 & 0.02 & 0.00 \\ 
  0.01 & 0.04 & 0.02 & 0.00 & 0.02 \\ 
  \end{psmallmatrix}$\\
   &\\ \hspace{2em}
$\mbox{\boldmath$\mu$}_1$  & $\begin{psmallmatrix}
  0.01 & 0.00 & 0.01 & 0.01 & 0.01 & 0.00 & 0.01 & 0.01 & 0.01 & 0.01 & 0.01 & 0.01 & 0.00 & 0.00 & 0.01 \\ 
  \end{psmallmatrix}$\\
 &\\ \hspace{2em}
$\mbox{\boldmath$\mu$}_2$  & $\begin{psmallmatrix}
  0.01 & 0.01 & 0.01 & 0.00 & 0.01 & 0.01 & 0.01 & 0.00 & 0.01 & 0.01 & 0.00 & 0.02 & 0.01 & 0.02 & 0.02 \\ 
  \end{psmallmatrix}$\\
 &\\ \hspace{2em}
$\text{diag}(\Wv_{1})$ & $\begin{psmallmatrix}
  0.00 & 0.00 & 0.00 & 0.00 & 0.00 & 0.00 & 0.01 & 0.00 & 0.00 & 0.01 & 0.01 & 0.00 & 0.00 & 0.00 & 0.00 \\ 
  \end{psmallmatrix}$\\
 &\\ \hspace{2em}
$\text{diag}(\Wv_{2})$ & $\begin{psmallmatrix}
  0.05 & 0.02 & 0.06 & 0.01 & 0.02 & 0.02 & 0.03 & 0.01 & 0.01 & 0.05 & 0.02 & 0.04 & 0.03 & 0.09 & 0.06 \\ 
  \end{psmallmatrix}$\\
 &\\ \hspace{2em}
$\text{diag}(\mbox{\boldmath$\Psi$}_1)$ & $\begin{psmallmatrix}
  0.01 & 0.00 & 0.01 & 0.00 & 0.00 & 0.01 & 0.04 & 0.00 & 0.00 & 0.00 & 0.01 & 0.00 & 0.00 & 0.00 & 0.00 \\ 
  \end{psmallmatrix}$\\
 &\\ \hspace{2em}
$\text{diag}(\mbox{\boldmath$\Psi$}_2)$ & $\begin{psmallmatrix}
  0.15 & 0.01 & 0.21 & 0.00 & 0.03 & 0.01 & 0.04 & 0.01 & 0.01 & 0.10 & 0.02 & 0.10 & 0.04 & 0.39 & 0.13 \\ 
  \end{psmallmatrix}$\\
\end{longtable}

\begin{longtable}{p{.125\textwidth}|p{.875\textwidth}}
%\scriptsize
%\centering
\caption{The average MSE of parameters when the correct model was selected (ICL) in Setting 3.}\label{simu1:setting3:MSE}\\
\hline 
Parameter&Mean squared Error\\
\hline
\\
\endfirsthead

\hline 
Parameter&Mean squared Error\\
\hline 
\\
\endhead
 &Continued on next page...\\ \hline
\endfoot
\hline
\endlastfoot
$\widetilde{\Bv}_1$& $\begin{psmallmatrix}
  0.32 & 0.50 & 0.33 & 0.14 & 0.64 & 0.53 & 0.13 & 0.41 & 0.45 & 0.22 \\ 
  0.06 & 0.04 & 0.05 & 0.14 & 0.20 & 0.02 & 0.05 & 0.09 & 0.27 & 0.13 \\ 
  0.10 & 0.10 & 0.18 & 0.06 & 0.06 & 0.19 & 0.37 & 0.36 & 0.53 & 0.04 \\ 
  0.09 & 0.07 & 0.27 & 0.06 & 0.36 & 0.12 & 0.16 & 0.03 & 0.08 & 0.36 \\ 
  0.05 & 0.22 & 0.05 & 0.23 & 0.12 & 0.22 & 0.02 & 0.15 & 0.03 & 0.22 \\ 
  0.27 & 0.25 & 0.33 & 0.14 & 0.18 & 0.18 & 0.08 & 0.15 & 0.17 & 0.18 \\ 
  0.07 & 0.08 & 0.11 & 0.12 & 0.10 & 0.72 & 0.04 & 0.30 & 0.03 & 0.02 \\ 
  0.14 & 0.17 & 0.11 & 0.27 & 0.11 & 0.06 & 0.06 & 0.11 & 0.02 & 0.19 \\ 
  0.15 & 0.87 & 0.03 & 0.08 & 0.05 & 0.40 & 0.01 & 0.04 & 0.31 & 0.74 \\ 
  0.27 & 0.09 & 0.06 & 0.04 & 0.12 & 0.06 & 0.09 & 0.34 & 0.01 & 0.19 \\ 
  0.13 & 0.34 & 0.69 & 0.05 & 0.24 & 0.08 & 0.05 & 0.05 & 0.21 & 0.48 \\ 
  0.15 & 0.15 & 0.21 & 0.51 & 0.35 & 0.14 & 0.07 & 0.04 & 0.04 & 0.12 \\ 
  0.24 & 0.07 & 0.10 & 0.15 & 0.03 & 0.15 & 0.80 & 0.31 & 0.02 & 0.06 \\ 
  0.11 & 0.09 & 0.05 & 0.03 & 0.04 & 0.24 & 0.23 & 0.26 & 0.05 & 0.09 \\ 
  0.13 & 0.16 & 0.45 & 0.14 & 0.06 & 0.45 & 0.03 & 0.04 & 0.29 & 1.00 \\ 
  0.24 & 0.08 & 0.03 & 0.03 & 0.21 & 0.42 & 0.15 & 0.17 & 0.02 & 0.11 \\ 
  \end{psmallmatrix}$ \\
     &\\ \hspace{2em}
  
$\widetilde{\Bv}_2$ & $\begin{psmallmatrix}
  1.29 & 0.90 & 2.16 & 0.95 & 0.88 & 2.04 & 0.49 & 0.56 & 0.70 & 1.76 \\ 
  0.14 & 0.44 & 0.10 & 1.21 & 0.21 & 0.33 & 0.18 & 0.14 & 0.78 & 0.70 \\ 
  0.37 & 0.18 & 0.27 & 0.11 & 0.55 & 0.64 & 0.31 & 1.88 & 0.82 & 0.83 \\ 
  0.18 & 0.13 & 1.23 & 0.06 & 0.80 & 1.12 & 1.49 & 0.03 & 0.19 & 0.84 \\ 
  0.06 & 1.00 & 0.13 & 0.34 & 0.58 & 0.54 & 0.78 & 2.21 & 0.29 & 1.15 \\ 
  0.85 & 0.51 & 0.12 & 0.09 & 0.19 & 0.11 & 0.07 & 0.16 & 0.22 & 0.25 \\ 
  0.08 & 1.19 & 0.09 & 0.41 & 0.10 & 2.37 & 0.08 & 0.78 & 0.05 & 0.74 \\ 
  0.42 & 0.61 & 0.07 & 0.25 & 0.08 & 0.43 & 0.05 & 1.05 & 0.30 & 0.33 \\ 
  0.58 & 0.95 & 1.03 & 1.13 & 0.05 & 1.63 & 0.04 & 0.07 & 0.38 & 1.18 \\ 
  0.78 & 0.17 & 0.81 & 0.19 & 0.24 & 0.25 & 0.10 & 1.34 & 0.55 & 1.64 \\ 
  0.53 & 0.26 & 0.86 & 0.67 & 0.30 & 1.29 & 0.13 & 0.29 & 0.29 & 0.77 \\ 
  0.30 & 0.87 & 0.73 & 0.45 & 0.62 & 0.29 & 0.07 & 0.35 & 0.57 & 0.73 \\ 
  0.33 & 0.11 & 0.23 & 0.23 & 1.52 & 1.36 & 6.06 & 0.42 & 0.07 & 0.62 \\ 
  0.07 & 0.52 & 0.17 & 0.20 & 0.14 & 0.27 & 0.23 & 0.29 & 0.29 & 0.18 \\ 
  0.07 & 0.19 & 2.56 & 0.10 & 1.20 & 1.24 & 0.06 & 0.11 & 0.24 & 3.79 \\ 
  0.22 & 1.19 & 0.16 & 0.16 & 0.22 & 0.60 & 1.15 & 0.29 & 0.36 & 0.25 \\ 
  \end{psmallmatrix}$ \\
     &\\ \hspace{2em}
  $\widetilde{\Bv}_3$ & $\begin{psmallmatrix}
  1.44 & 0.91 & 2.14 & 0.90 & 0.51 & 1.51 & 0.68 & 0.52 & 0.53 & 1.79 \\ 
  0.17 & 0.49 & 0.11 & 1.00 & 0.15 & 0.34 & 0.15 & 0.09 & 0.53 & 0.59 \\ 
  0.26 & 0.06 & 0.12 & 0.07 & 0.50 & 0.45 & 0.14 & 1.43 & 0.37 & 0.89 \\ 
  0.13 & 0.15 & 0.91 & 0.09 & 0.55 & 1.12 & 1.17 & 0.05 & 0.11 & 0.44 \\ 
  0.04 & 0.79 & 0.15 & 0.18 & 0.49 & 0.29 & 0.79 & 2.02 & 0.32 & 0.89 \\ 
  0.81 & 0.42 & 0.08 & 0.08 & 0.13 & 0.09 & 0.07 & 0.13 & 0.12 & 0.11 \\ 
  0.16 & 1.17 & 0.13 & 0.31 & 0.09 & 1.69 & 0.10 & 0.77 & 0.07 & 0.73 \\ 
  0.33 & 0.61 & 0.06 & 0.25 & 0.08 & 0.37 & 0.06 & 0.99 & 0.29 & 0.19 \\ 
  0.45 & 0.35 & 1.01 & 1.02 & 0.02 & 1.18 & 0.04 & 0.07 & 0.15 & 0.55 \\ 
  0.56 & 0.10 & 0.75 & 0.16 & 0.15 & 0.21 & 0.04 & 0.96 & 0.53 & 1.72 \\ 
  0.48 & 0.13 & 0.41 & 0.79 & 0.27 & 1.42 & 0.09 & 0.27 & 0.14 & 0.40 \\ 
  0.28 & 0.69 & 0.56 & 0.33 & 0.38 & 0.27 & 0.06 & 0.42 & 0.63 & 0.62 \\ 
  0.12 & 0.07 & 0.16 & 0.14 & 1.61 & 1.09 & 5.11 & 0.17 & 0.06 & 0.55 \\ 
  0.04 & 0.40 & 0.15 & 0.22 & 0.15 & 0.17 & 0.08 & 0.17 & 0.26 & 0.15 \\ 
  0.07 & 0.14 & 2.11 & 0.08 & 1.16 & 0.94 & 0.07 & 0.10 & 0.08 & 2.71 \\ 
  0.06 & 1.11 & 0.16 & 0.19 & 0.07 & 0.27 & 1.02 & 0.21 & 0.30 & 0.17 \\ 
  \end{psmallmatrix}$ \\
     &\\ \hspace{2em}
$\mbox{\boldmath$\Sigma$}_{\ev_1}$ &$ \begin{psmallmatrix}
  0.52 & 0.23 & 0.50 & 0.12 & 0.12 & 0.12 & 0.08 & 0.09 & 0.05 & 0.04 \\ 
  0.23 & 0.19 & 0.14 & 0.11 & 0.10 & 0.08 & 0.04 & 0.06 & 0.03 & 0.03 \\ 
  0.50 & 0.14 & 0.34 & 0.12 & 0.11 & 0.17 & 0.27 & 0.19 & 0.02 & 0.10 \\ 
  0.12 & 0.11 & 0.12 & 0.11 & 0.10 & 0.13 & 0.05 & 0.06 & 0.15 & 0.02 \\ 
  0.12 & 0.10 & 0.11 & 0.10 & 0.11 & 0.10 & 0.06 & 0.15 & 0.04 & 0.06 \\ 
  0.12 & 0.08 & 0.17 & 0.13 & 0.10 & 0.13 & 0.06 & 0.19 & 0.03 & 0.10 \\ 
  0.08 & 0.04 & 0.27 & 0.05 & 0.06 & 0.06 & 0.07 & 0.09 & 0.03 & 0.01 \\ 
  0.09 & 0.06 & 0.19 & 0.06 & 0.15 & 0.19 & 0.09 & 0.11 & 0.21 & 0.02 \\ 
  0.05 & 0.03 & 0.02 & 0.15 & 0.04 & 0.03 & 0.03 & 0.21 & 0.06 & 0.04 \\ 
  0.04 & 0.03 & 0.10 & 0.02 & 0.06 & 0.10 & 0.01 & 0.02 & 0.04 & 0.02 \\ 
  \end{psmallmatrix}$\\
     &\\ \hspace{2em}
$\mbox{\boldmath$\Sigma$}_{\ev_2}$ & $\begin{psmallmatrix}
  1.34 & 0.88 & 0.48 & 1.00 & 1.03 & 1.15 & 0.43 & 0.43 & 0.26 & 0.05 \\ 
  0.88 & 0.51 & 0.89 & 0.24 & 0.87 & 0.17 & 0.05 & 0.33 & 0.23 & 0.11 \\ 
  0.48 & 0.89 & 1.04 & 0.26 & 0.57 & 0.08 & 0.17 & 0.10 & 0.02 & 0.06 \\ 
  1.00 & 0.24 & 0.26 & 0.24 & 0.32 & 0.28 & 0.13 & 0.37 & 0.63 & 0.12 \\ 
  1.03 & 0.87 & 0.57 & 0.32 & 1.15 & 0.77 & 0.43 & 0.36 & 0.08 & 0.17 \\ 
  1.15 & 0.17 & 0.08 & 0.28 & 0.77 & 0.23 & 0.31 & 0.42 & 0.15 & 0.28 \\ 
  0.43 & 0.05 & 0.17 & 0.13 & 0.43 & 0.31 & 0.20 & 0.31 & 0.26 & 0.05 \\ 
  0.43 & 0.33 & 0.10 & 0.37 & 0.36 & 0.42 & 0.31 & 0.28 & 0.50 & 0.03 \\ 
  0.26 & 0.23 & 0.02 & 0.63 & 0.08 & 0.15 & 0.26 & 0.50 & 0.31 & 0.22 \\ 
  0.05 & 0.11 & 0.06 & 0.12 & 0.17 & 0.28 & 0.05 & 0.03 & 0.22 & 0.02 \\ 
  \end{psmallmatrix}$\\
     &\\ \hspace{2em}
  $\mbox{\boldmath$\Sigma$}_{\ev_3}$ & $\begin{psmallmatrix}
  0.59 & 0.45 & 0.32 & 0.59 & 0.57 & 0.61 & 0.24 & 0.20 & 0.11 & 0.03 \\ 
  0.45 & 0.30 & 0.42 & 0.19 & 0.45 & 0.13 & 0.10 & 0.22 & 0.16 & 0.10 \\ 
  0.32 & 0.42 & 0.55 & 0.20 & 0.31 & 0.08 & 0.11 & 0.06 & 0.02 & 0.03 \\ 
  0.59 & 0.19 & 0.20 & 0.22 & 0.11 & 0.17 & 0.07 & 0.20 & 0.33 & 0.10 \\ 
  0.57 & 0.45 & 0.31 & 0.11 & 0.53 & 0.33 & 0.23 & 0.14 & 0.04 & 0.08 \\ 
  0.61 & 0.13 & 0.08 & 0.17 & 0.33 & 0.18 & 0.14 & 0.14 & 0.06 & 0.14 \\ 
  0.24 & 0.10 & 0.11 & 0.07 & 0.23 & 0.14 & 0.09 & 0.14 & 0.17 & 0.07 \\ 
  0.20 & 0.22 & 0.06 & 0.20 & 0.14 & 0.14 & 0.14 & 0.12 & 0.16 & 0.02 \\ 
  0.11 & 0.16 & 0.02 & 0.33 & 0.04 & 0.06 & 0.17 & 0.16 & 0.10 & 0.12 \\ 
  0.03 & 0.10 & 0.03 & 0.10 & 0.08 & 0.14 & 0.07 & 0.02 & 0.12 & 0.02 \\ 
  \end{psmallmatrix}$\\
     &\\ \hspace{2em}
$\mbox{\boldmath$\mu$}_1$  & $\begin{psmallmatrix}
  0.78 & 0.15 & 0.09 & 0.27 & 0.62 & 0.51 & 0.12 & 0.39 & 0.17 & 0.74 & 0.13 & 0.13 & 0.61 & 0.07 & 0.20 \\ 
  \end{psmallmatrix}$\\
   &\\ \hspace{2em}
$\mbox{\boldmath$\mu$}_2$  & $\begin{psmallmatrix}
  0.97 & 0.84 & 1.27 & 0.48 & 0.76 & 1.46 & 1.42 & 1.14 & 1.03 & 0.83 & 0.59 & 1.37 & 0.64 & 1.20 & 1.00 \\ 
  \end{psmallmatrix}$\\
   &\\ \hspace{2em}
$\mbox{\boldmath$\mu$}_3$  & $\begin{psmallmatrix}
  0.66 & 0.86 & 1.30 & 0.40 & 0.53 & 1.37 & 1.50 & 1.07 & 1.07 & 0.55 & 0.60 & 1.49 & 0.41 & 1.27 & 1.04 \\ 
  \end{psmallmatrix}$\\
   &\\ \hspace{2em}
$\text{diag}(\Wv_{1})$ & $\begin{psmallmatrix}
  0.02 & 0.04 & 0.02 & 0.06 & 0.03 & 0.06 & 0.01 & 0.21 & 0.13 & 0.01 & 0.07 & 0.12 & 0.02 & 0.20 & 0.02 \\ 
  \end{psmallmatrix}$\\
   &\\ \hspace{2em}
$\text{diag}(\Wv_{2})$ & $\begin{psmallmatrix}
  0.07 & 0.17 & 0.06 & 0.25 & 0.04 & 0.17 & 0.02 & 0.41 & 0.27 & 0.03 & 0.15 & 0.30 & 0.12 & 0.39 & 0.13 \\ 
  \end{psmallmatrix}$\\
   &\\ \hspace{2em}
$\text{diag}(\Wv_{3})$ & $\begin{psmallmatrix}
  0.05 & 0.14 & 0.05 & 0.18 & 0.03 & 0.11 & 0.01 & 0.23 & 0.17 & 0.01 & 0.12 & 0.15 & 0.11 & 0.19 & 0.09 \\ 
  \end{psmallmatrix}$\\
   &\\ \hspace{2em}
$\text{diag}(\mbox{\boldmath$\Psi$}_1)$ & $\begin{psmallmatrix}
  0.05 & 0.01 & 0.10 & 0.02 & 0.03 & 0.02 & 0.01 & 0.01 & 0.00 & 0.01 & 0.02 & 0.04 & 0.01 & 0.00 & 0.03 \\ 
  \end{psmallmatrix}$\\
   &\\ \hspace{2em}
$\text{diag}(\mbox{\boldmath$\Psi$}_2)$ & $\begin{psmallmatrix}
  0.04 & 0.17 & 0.17 & 0.05 & 0.06 & 0.21 & 0.03 & 0.06 & 0.10 & 0.02 & 0.06 & 0.15 & 0.02 & 0.04 & 0.07 \\ 
  \end{psmallmatrix}$\\
   &\\ \hspace{2em}
$\text{diag}(\mbox{\boldmath$\Psi$}_3)$ & $\begin{psmallmatrix}
  0.03 & 0.16 & 0.09 & 0.05 & 0.07 & 0.17 & 0.01 & 0.04 & 0.08 & 0.01 & 0.04 & 0.09 & 0.02 & 0.03 & 0.06 \\ 
  \end{psmallmatrix}$\\
\end{longtable}

\end{appendices}
\end{document}